\documentclass[aps,preprint,showpacs,superscriptaddress,a4paper,nofootinbib]{revtex4}
\pdfoutput=1

\usepackage{graphicx, color}
\usepackage{epstopdf}           

\usepackage{amsmath}
\usepackage{amssymb}
\usepackage{latexsym}
\usepackage{times}

\newcommand{\scri}{\ensuremath{\mathcal{J}^+}}

\newcommand{\ie}{\textit{i.e.}}
\newcommand{\eg}{\textit{e.g.}}
\DeclareMathOperator{\p}{\ensuremath{\partial}}

\def\be{\begin{equation}}
\def\ee{\end{equation}}
\def\bea{\begin{eqnarray}}
\def\eea{\end{eqnarray}}

\begin{document}

\title{Characteristic extraction in numerical relativity: binary black
  hole merger waveforms at null infinity}

\author{C. Reisswig}
\affiliation{
Max-Planck-Institut f\" ur Gravitationsphysik, Albert-Einstein-Institut,
14476 Golm, Germany
}

\author{N.~T. Bishop}
\affiliation{
Department of Mathematics, Rhodes University, Grahamstown 6140, South
Africa
}

\author{D. Pollney}
\affiliation{
Max-Planck-Institut f\" ur Gravitationsphysik, Albert-Einstein-Institut,
14476 Golm, Germany
}

\author{B. Szil\'{a}gyi}
\affiliation{
Theoretical Astrophysics, California Institute of Technology,
Pasadena, CA 91125, USA
}

\begin{abstract}
The accurate modeling of gravitational radiation is a key issue for
gravitational wave astronomy.
As simulation codes reach higher accuracy, systematic errors
inherent in current numerical relativity wave-extraction methods become evident, and may lead to a wrong astrophysical interpretation of the data.
In this paper, we give a detailed description of the
\emph{Cauchy-characteristic extraction} technique applied to binary black hole inspiral and merger
evolutions to obtain gravitational waveforms that are defined unambiguously, that is, at future null infinity.
By this method we remove finite-radius approximations and the need
to extrapolate data from the near zone. Further, we demonstrate that
the method is free of gauge effects and thus is affected only by numerical error.
Various consistency checks reveal that energy and angular momentum are conserved to high precision and agree very well with extrapolated
data. In addition, we revisit the computation of the gravitational recoil and find that finite radius extrapolation very well approximates the result at $\scri$.
However, the (non-convergent) systematic differences to extrapolated data are of the same order of magnitude as the (convergent) discretisation error of the Cauchy evolution hence
highlighting the need for correct wave-extraction.
\end{abstract}


\maketitle

\section{Introduction}

The computation of gravitational radiation
from black hole merger events has attracted considerable attention,
since the pioneering work by Smarr and
collaborators~\cite{Smarr75, Smarr76, Smarr78b}.  With the advent of
ground-based laser interferometric gravitational wave detectors, as
well as the prospect of the Laser Interferometer Space Antenna (LISA),
interest in the problem has increased considerably in recent years.  The measurement
of gravitational waves is expected to provide an important probe of
strong-field nonlinear gravity, thereby enabling empirical testing of this
regime, as well as to provide direct observations of processes of fundamental
interest in astrophysics.  The sensitivities of LISA, and of the
upcoming advanced ground-based detectors AdLIGO and AdVirgo, are high enough
that even an error in aspects of the waveform calculation of $0.1\%$
could lead to an incorrect interpretation of
the astrophysical properties of a source, or of a test of
general relativity~\cite{Reisswig:2009us}.

Nowadays, there are
several codes that can produce a stable and convergent simulation of a
black hole spacetime \cite{Pretorius:2005gq, Baker:2005vv, Campanelli:2005dd,
Pollney:2007ss,  Scheel:2008rj, Pollney:2009yz, Vaishnav:2007nm, Liebling-2002:nonlinear-sigma-criticality-via-3D-AMR, Bruegmann:2006at, Husa:2007rh},
and of particular topical relevance
is the construction of long and accurate
waveforms which can be used to analyze distinguishable features within
the signals~\cite{Reisswig:2009vc}, 
and also
to construct a family of
templates with the goal of improving gravitational wave
detection algorithms. Here the requirements are particularly
challenging for numerical simulations, requiring waveforms which are
accurate in phase and amplitude over many cycles to allow for an
unambiguous matching to post-Newtonian waveforms at large
separation. Some recent studies have shown very promising results in
this direction for particular binary black hole
models~\cite{Baker:2006ha, Buonanno06imr, Hannam:2007ik,
  Hannam:2007wf, Damour:2007vq, Buonanno:2007pf, Damour:2008te,
  Buonanno:2009qa, Boyle:2007ft, Ajith:2007qp,Ajith:2007kx,Ajith:2007xh,Ajith:2009bn}. 

However, a particular difficulty arises from the fact that
gravitational energy cannot be defined locally in general relativity,
and is well-defined only at future null infinity, $\scri$.  Since
standard numerical evolutions are carried out on finite domains, there
is a systematic error caused by estimating the gravitational radiation
from fields on a world-tube at finite radius and the uncertainty in
how it relates to measurement at $\scri$~\cite{Kocsis:2007}.  Even
though this error may be small, the expected sensitivity of upcoming
detectors means that it is important to obtain an accurate result.

A rigorous formalism for the global measurement of gravitational
energy at $\scri$ has been in place since the pioneering work of
Bondi, Penrose and collaborators in the 1960s~\cite{Bondi62,
Penrose:1963}; and subsequently, techniques for calculating gravitational 
radiation at $\scri$ have been developed. The idea which we pursue here
is to combine a standard Cauchy or ``3~+~1'' numerical relativity code with a
characteristic code~\cite{Bishop93} in order to transport the wave
information
to $\scri$ using the full Einstein equations. This method involves a Cauchy computation
within a domain bounded by a world-tube, $\Gamma$, together with a
characteristic computation with inner boundary $\Gamma$. If the
data is consistently passed back and forth between the evolution schemes at
$\Gamma$ during the evolution, the method is known as 
\emph{Cauchy-characteristic matching} (CCM). Given astrophysical initial data,
such a method has only discretisation error~\cite{Bishop96}, and a
complete mathematical specification has been
developed~\cite{Bishop98b}. However, efforts to implement CCM 
encountered stability problems at the interface, and in the general
case have not been successful. An alternative approach is
\emph{Cauchy-characteristic extraction} (CCE), or characteristic extraction.
In this case, an independent Cauchy computation is carried out initially,
with its usual timelike artificial outer boundary and which
knows nothing about a characteristic code. Data on the world-tube
$\Gamma$ is stored, and then used to provide inner boundary data for a
characteristic code -- but not vice-versa. Initial implementations
of CCE~\cite{Babiuc:2005pg,Babiuc:2009} have shown the effectiveness
of the method for a number of test problems.

In this paper we describe the implementation of a CCE
code, and present results
obtained from the code for the real astrophysical problem of the
inspiral and merger waveform of two
black holes. The gravitational waveforms are calculated at
$\scri$, and are thus, for this case, the first waveforms that are
unambiguous in the sense of being free of gauge 
and finite-radius effects. 
Further, the code is general purpose, in that it is
independent of the details of the Cauchy code: it runs as a post-processing
operation, taking as input only
the required geometrical data on a world-tube. Thus it can be used
in combination with other Cauchy codes, and applied to
other astrophysical problems.

The next section summarizes notation and results
needed from other work. Then, Sec.~\ref{s-imp} describes in some detail
our implementation of characteristic extraction. Sec.~\ref{s-test}
presents results of testing the code against analytic solutions. The
computation, including results, of the inspiral and merger of two
black holes is described in Sec.~\ref{sec:results}. Some technical
details of the characteristic and extraction code parameters are
described in Sec.~\ref{s-param}.

\section{Review of results needed from other work}
\label{s-rev}

\subsection{The Bondi-Sachs metric}
The formalism for the numerical evolution of Einstein's equations, in
null cone  coordinates, is well known~\cite{Bondi62,Isaacson83,Bishop96,Bishop97b,Gomez01,Bishop99}.
For the sake of completeness, 
we give a summary of those aspects of the formalism that will be used here.
We start with coordinates based upon a family of outgoing null
hypersurfaces.
We let $u$ label these hypersurfaces, $x^A$ $(A=2,3)$, label
the null rays, and $r$ is a surface area coordinate. In the resulting
$x^\alpha=(u,r,x^A)$ coordinates, the metric takes the Bondi-Sachs
form~\cite{Bondi62,Sachs62}
\begin{eqnarray}
 ds^2  &=&  -\left(e^{2\beta}(1 + W_c r) -r^2h_{AB}U^AU^B\right)du^2
\nonumber \\
        & &- 2e^{2\beta}dudr -2r^2 h_{AB}U^Bdudx^A 
        +  r^2h_{AB}dx^Adx^B,
\label{eq:bmet}
\end{eqnarray}
where $h^{AB}h_{BC}=\delta^A_C$ and
$det(h_{AB})=\det(q_{AB})$, with $q_{AB}$ a metric representing a unit
2-sphere embedded in flat Euclidean 3-space; $W_c$ is
a normalized variable used in the code, related to the usual Bondi-Sachs
variable $V$ by $V=r+W_c r^2$. As discussed in more detail below, we
represent $q_{AB}$ by means of a complex dyad $q_A$. Then,
for an arbitrary Bondi-Sachs metric,
$h_{AB}$ can be represented by its dyad component
\begin{equation}
J=h_{AB}q^Aq^B/2,
\end{equation}
with the spherically symmetric case characterized by $J=0$.
We also introduce the fields
\begin{equation}
K=\sqrt{1+J \bar{J}},\qquad
U=U^Aq_A,
\end{equation}
as well as the (complex differential) eth operators $\eth$ and $\bar \eth$
~\cite{Gomez97}.

The 10 Einstein equations $R_{\alpha\beta}=8\pi(T_{\alpha\beta}
-\frac{1}{2}g_{\alpha\beta}T)$ are classified as: (i) hypersurface
equations -- $R_{11},q^AR_{1A},h^{AB}R_{AB}$ -- forming a hierarchical
set for $\beta,U$ and $W_c$; (ii) evolution equation $q^Aq^B R_{AB}$ for
$J$; and (iii) constraints $R_{0\alpha}$. An evolution problem is normally
formulated in the region of spacetime between a timelike or null
world-tube, $\Gamma$, and future null infinity ($\scri$), with
(free) initial data,
$J$, given on $u=0$, and with boundary data for $\beta,U,W_c,J$ satisfying
the constraints given on the inner world-tube. The inclusion of $\scri$
in the computational grid is made possible by compactifying the radial
coordinate, $r$, for instance by means of a transformation of the form
\begin{equation}
r \rightarrow x=\frac{r}{r+r_{\rm wt}}\,.
\label{eq:compact}
\end{equation}
In characteristic coordinates (but not in general), the Einstein equations
remain regular at $\scri$ under such a transformation.


\subsection{The spin-weighted formalism and the $\eth$ operator}
A complex dyad $q_A$ is a 2-vector whose real and imaginary parts are
unit vectors that are orthogonal to each other. Further, $q_A$ represents
the metric, and has the properties
\begin{equation}
q^A q_A = 0, \qquad q^A \bar q_A = 2,\qquad
q_{AB}=\frac{1}{2}(q_A\bar{q}_B+\bar{q}_A q_B).
\label{eq:diad.norm.def}
\end{equation}
Note that $q_A$ is not unique, up to a unitary factor: if $q_A$ represents
a given 2-metric, then so does $q^\prime_A=e^{i\alpha}q_A$. Thus,
considerations of simplicity are used in deciding the precise form of
dyad to represent a particular 2-metric. The dyad commonly used to
represent the unit sphere metric in the stereographic coordinates used here, is
\begin{equation}
ds^2=\frac{4(dq^2+dp^2)}{(1+q^2+p^2)^2},\qquad q_A=\frac{2}{1+q^2+p^2}(1,i).
\end{equation}

Having defined a dyad, we may construct complex quantities representing
all manner of tensorial objects, for example $X_1=T_A q^A$,
$X_2=T^{AB} q_A \bar{q}_B$, $X_3=T^{AB}_C \bar{q}_A\bar{q}_B\bar{q}^C$.
Each object has no free indices, and has associated with it a spin-weight
$s$ defined as the number of $q$ factors less the number of $\bar{q}$
factors in its definition. For example, $s(X_1)=1,s(X_2)=0,s(X_3)=-3$,
and, in general, $s(X)=-s(\bar{X})$.
We define derivative operators $\eth$ and $\bar{\eth}$ acting on
a quantity $V$ with spin-weight $s$
\begin{equation}
\eth V=q^A \partial_A V + s \Gamma V,\qquad
\bar{\eth} V=\bar{q}^A \partial_A V - s \bar{\Gamma} V
\end{equation}
where the spin-weights of $\eth V$ and $\bar{\eth} V$ are $s+1$ and $s-1$,
respectively, and where
\begin{equation}
\Gamma=-\frac{1}{2} q^A\bar{q}^B\nabla_A q_B.
\label{e-G}
\end{equation}
In the case of stereographic coordinates, $\Gamma=q+ip$.

The spin-weights of the quantities used in the Bondi-Sachs metric are
\begin{equation}
s(W_c)=s(\beta)=0,\quad s(J)=2,\quad s(\bar{J})=-2,\quad s(K)=0,\quad s(U)=1,\quad s(\bar{U})=-1.
\end{equation}

We will be using spin-weighted spherical
harmonics~\cite{Newman-Penrose-1966,Goldberg:1967}
${}_s Y_{\ell m}$ (the suffix ${}_s$ denotes the spin-weight),
using the formalism described in~\cite{Zlochower03}, and note that this
reference gives explicit formulas for the ${}_s Y_{\ell m}$ in stereographic
coordinates. In the case $s=0$, the $s$ will be omitted, i.e.
$Y_{\ell m}={}_0 Y_{\ell m}$.
Note that the effect of the $\eth$ operator acting on $Y_{\ell m}$ is
\begin{equation}
\eth Y_{\ell m}=\sqrt{\ell(\ell+1)}\;{}_1Y_{\ell m}, \qquad
\eth^2 Y_{\ell m}=\sqrt{(\ell -1)\ell(\ell+1)(\ell+2)}\;{}_2Y_{\ell m}.
\end{equation}

\subsection{Gravitational radiation: News and $\psi_4$}
The mathematical theory relating metric quantities at future null infinity,
${\mathcal J}^+$, to gravitational radiation, and using the present formalism,
is given in~\cite{Bishop97b,Bishop03,Babiuc:2009}. In the original work of Bondi
{\it et al.}~\cite{Bondi62}, it was possible to assume that the coordinates
at ${\mathcal J}^+$ were such that $\beta=J=0$ there, and in that case the
 gravitational news takes the very simple form
\begin{equation}
   N=\frac{1}{2} \partial_u \partial_\rho J
\end{equation}
(where we have written the radial derivative in terms of $\rho :=
r^{-1}$ rather than the usual notation, $\ell$, to
avoid confusion with the spherical harmonic index $\ell$). However, the
coordinates used in the characteristic code are fixed at the inner
boundary $\Gamma$, and in general the Bondi gauge conditions are not satisfied.
Previous work has presented the formalism for calculating, in a general gauge,
the gravitational news~\cite{Bishop97b}, as well as $\psi_4$~\cite{Babiuc:2009}.

Geometrically, the Bondi gauge condition $J=0$ means that, for large $r$, a
2-surface of constant $(u,r)$ is spherical rather than being, for example,
an ellipsoid; and this condition can be expressed algebraically by saying that
the 2-surface has constant curvature. An expression for the news, in a general
gauge, must take (implicit) account of the transformation to Bondi gauge
coordinates,
\begin{equation}
r\rightarrow r_{[B]}= \omega(u,x^A) r,\qquad
x^A \rightarrow x_{[B]}^A=x_{[B]}^A(u,x^A).
\end{equation}
The constant curvature condition leads to the factor $\omega$ satisfying
\begin{equation}
2K-\eth\bar{\eth} K +\frac{1}{2}\left(\bar{\eth}^2J +\eth^2\bar{J}\right)
+\frac{1}{4K}\left( \bar{\eth}\bar{J}\eth J - \bar{\eth}J \eth \bar{J}\right)
=2\left(\omega^2+h^{AB}D_A D_B \log\omega \right)
\end{equation}
with all metric quantities evaluated at ${\mathcal J}^+$. Alternatively, we can
initialize $J=0$ at $u=0$, so that $\omega=1$ at $u=0$, and then evolve
$\omega$ by
\begin{equation}
\partial_u \omega=-\frac{1}{2}( \bar{U}\eth\omega +U\bar{\eth}\omega)
                  -\frac{\omega}{4} (\eth \bar{U}+\bar{\eth}U).
\end{equation}

The Bondi gauge condition $\beta=0$ at ${\mathcal J}^+$ means that, for large
$r$, coordinate time is the same as proper time, and the implementation of
this condition is straightforward.

While the transformation to Bondi gauge coordinates can be done
explicitly~\cite{Bishop03}, in the present code it is done implicitly, leading to
an expression for the news in terms of the code metric variables and coordinates.
The formula is long and complicated~\cite{Bishop97b} and is not repeated here.
However, in the linearized case, the news function is much
simpler~\cite{Bishop-2005b}. The results
presented later are approximately in the linearized regime, and so it can be
instructive to examine this case (even though the code is fully nonlinear).
The linearized formula for the news is
\begin{equation}
   N=\frac{1}{2} \partial_u \partial_\rho J
      +\frac{1}{2} \eth^2(\omega +2 \beta).
\label{eq:linN}
\end{equation}
In this case $\omega$ is simply related to $J$ at ${\mathcal J}^+$, if
the metric is decomposed into spin-weighted spherical harmonics,
\begin{equation}
J=\sum_{\ell m} J_{\ell m}\; {}_2Y_{\ell m}, \;\;
\beta=\sum_{\ell m} \beta_{\ell m}\; Y_{\ell m},
\end{equation}
Then the news is
\begin{equation}
N=\sum_{\ell m}  N_{\ell m}\; {}_2Y_{\ell m} = \sum_{\ell m} 
\left(\frac{1}{2} \partial_u \partial_\rho J_{\ell m}
 - \frac{\ell(\ell+1)}{4}J_{\ell m}
 +\beta_{\ell m} \right) \sqrt{(\ell-1)\ell(\ell+1)(\ell+2)}\;{}_2Y_{\ell m}.
\label{e-N}
\end{equation}

In characteristic work, it is conventional to work with a quantity $\Psi_4$
rather than the usual $\psi_4$ used in current Cauchy codes (see \eg \cite{Pollney:2009yz}). 
The relationship is
\begin{equation}
\Psi_4=-(1/2)\bar \psi_4\,,
\label{eq:psiPsi}
\end{equation}
and we will usually transform the characteristic $\Psi_4$ to the standard $\psi_4$.
The formulas for $\Psi_4$ are long and complicated and are not given here,
but in the linearized case, $\Psi_4$ can be expressed in a simple way in terms of
metric quantities at ${\mathcal J}^+$~\cite{Babiuc:2009} (in which the
notation $\Psi$ rather than $\Psi_4$ is used) 
\begin{equation}
      \Psi_4=\frac{1}{2}\partial_u^2 \partial_\rho J -\frac{1}{2}\partial_u J
      -\frac{1}{2}\eth U -\frac{1}{8} \eth^2( \eth \bar U +\bar \eth U)
       + \partial_u \eth^2 \beta.
\label{eq:linPsi}
\end{equation}
It can be shown that the linearized Einstein equations lead to the
relation $\Psi_4=\partial_u N$.

\subsection{Characteristic extraction}
\label{sec:char-extr}
We refer to~\cite{Bishop98b} (see also~\cite{Szilagyi00,Babiuc:2005pg})
for a full description of the characteristic extraction process.
While characteristic extraction appears to be only a coordinate transformation,
it is actually rather more complicated. The difficulty is that Bondi-Sachs
coordinates use a surface area radial coordinate, and this coordinate can be
constructed only once the angular coefficients of the metric have been found.
Thus the construction proceeds in two stages.

\begin{figure}
\includegraphics[scale=0.7]{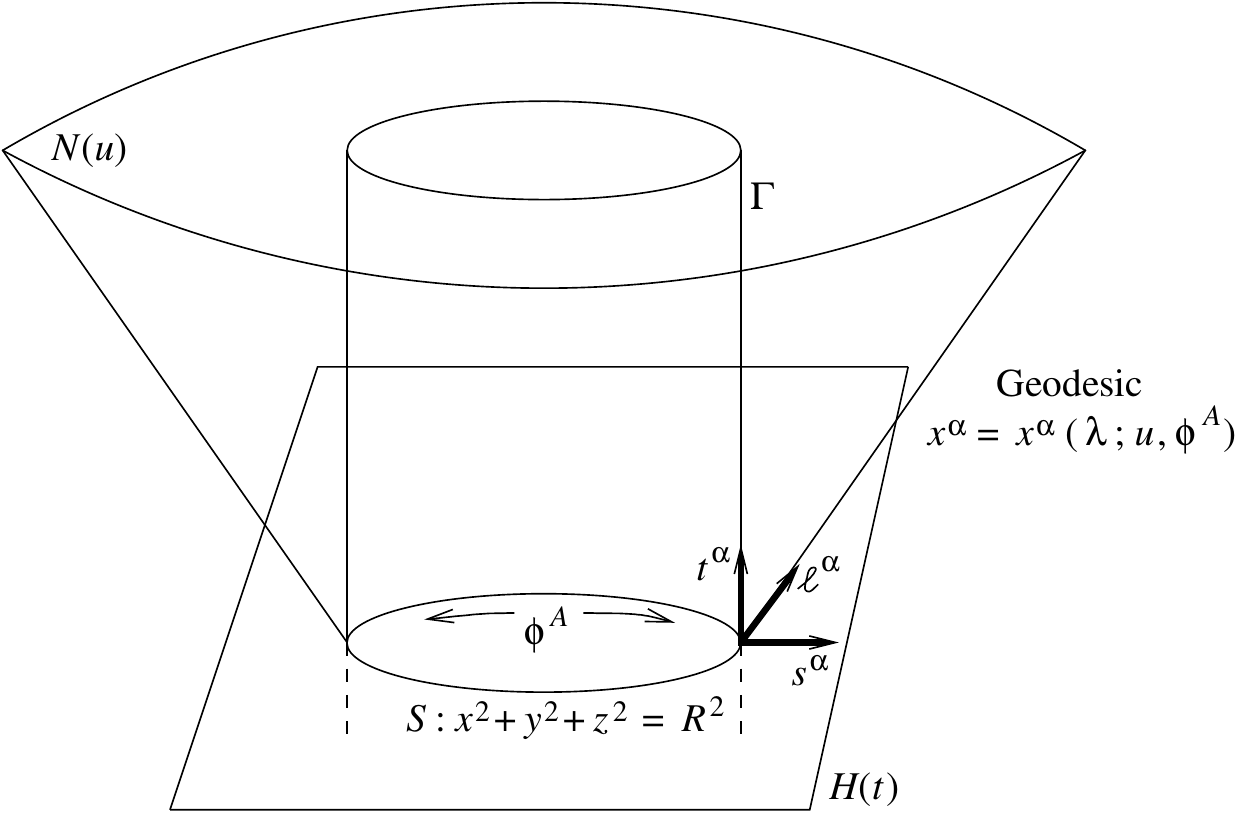}
\caption{Schematic illustration of the (first stage) construction of
characteristic coordinates and metric.}
\label{f-cce-idea}
\end{figure}

In the first stage, we use an affine coordinate $\lambda$ in the radial
direction, and find the transformed metric and its first ${}_{,\lambda}$
derivatives at angular grid-points of the extraction world-tube $\Gamma$.
The process is illustrated in Fig.~\ref{f-cce-idea}, and can be
summarized as follows:
\begin{itemize}
\item Define a world-tube $\Gamma$ by $x^2+y^2+z^2=R^2$ with $R$ constant,
and induce angular coordinates $\phi^A$ on $\Gamma$ as though in Euclidean space.
\item Let H be a hypersurface of constant $t$, and define the 2-surface
$S=H \cap \Gamma$.
\item Let $t^\alpha$ be a unit normal to $H$, and let
$s^\alpha$ be normal to $S$ in $H$.
\item Construct outgoing null vectors
$\ell^\alpha=t^\alpha+s^\alpha$, and then outgoing null geodesics
in direction $\ell^\alpha$ with affine parameter $\lambda$.
\item The union
of such geodesics is the null cone $N$, labeled by $u$ being the Cauchy time $t$
where $N$ meets $\Gamma$.
\item Construct the Jacobian for the coordinate transformation
$(t,x,y,z)\rightarrow(u,\lambda,\phi^A)$.
\item Find the transformed metric at the angular grid-points of $\Gamma$.
\item Find the first ${}_{,\lambda}$ derivatives of the transformed metric
at the angular grid-points of $\Gamma$.
\end{itemize}

In the second stage, we make the transformation to a surface area coordinate $r$.
The difficulty here is that, in general, $r$ is not constant on $\Gamma$. Thus
in order to set data on an inner world-tube of the characteristic grid, we need
the metric quantities on the world-tube, as well as their first derivatives off
the world-tube. The process can be summarized as follows:
\begin{itemize}
\item Make the coordinate transformation
$\lambda\rightarrow r=r(u,\lambda,\phi^A)$
with $r$ defined by the condition that it is a surface area coordinate.
\item Find $r$ and $r_{,\lambda}$ at the grid points of $\Gamma$.
\item Find the metric and its first ${}_{,\lambda}$ derivatives at the grid
points of $\Gamma$.
\item Find $J$, $U$, $W_c$ and $\beta$ at the grid points of $\Gamma$.
\item Using $\partial_r=\partial_\lambda /r_{,\lambda}$, find $J_{,r}$,
$U_{,r}$, $W_{c,r}$ and $\beta_{,r}$ at the grid points of $\Gamma$.
\item Set $J$, $U$, $U_{,r}$, $W_c$ and $\beta$ on an inner world-tube of the
characteristic grid.
\end{itemize}
The value used for $U_{,r}$ is second-order accurate on the world-tube
$\Gamma$, and it is this value which is used at the inner world-tube of
the characteristic grid, so, in general, it is only first-order accurate.
However, in practice, the overall performance of the code remains second-order
accurate.

\subsection{The Cauchy evolution code and finite radius measurements}

A fundamental problem with early attempts to implement CCM was the
sensitivity of the 3+1 (Cauchy) implementation to boundary
perturbations. In recent years, however, a number of strongly
hyperbolic formulations have been developed and proven to be
numerically robust for binary black hole evolutions using artificial
outer boundary conditions
We evolve the spacetime in the neighborhood of the binary black holes
using a variant of the ``BSSNOK'' evolution system~\cite{Nakamura87,
  Shibata95, Baumgarte99}, according to the implementation described
in~\cite{Pollney:2009yz}. This formulation uses a conformally
transformed metric, $\tilde{\gamma}_{ab}$, and (traceless) extrinsic
curvature, $\tilde{A}_{ab}$, as evolution variables, where the
conformal transformation is defined by
\begin{equation}
  \phi  = \frac{1}{12} \ln\det\gamma_{ab},\qquad
  \tilde{\gamma}_{ab} := e^{-4\phi}\gamma_{ab}.
\end{equation}
Instead of evolving $\phi$ directly, we use the concomitant
\begin{equation}
  \hat{\phi}_6 := (\det\gamma_{ab})^{-1/6} = e^{-2\phi},
\end{equation}
following the suggestion of~\cite{Marronetti:2007wz}.

Initial data for binary black holes is determined using the
\emph{puncture} method~\cite{Brandt97b}, and corresponds to a
conformally flat solution of the Hamiltonian constraint with the
extrinsic curvature specified according to Bowen and
York~\cite{Bowen80}. Parameters for quasi-circular orbits starting
from a given separation are determined by a post-Newtonian
estimate~\cite{Husa:2007rh}, resulting in a low eccentricity
($e\simeq0.04$) inspiral for the equal-mass non-spinning binary
considered below.

The lapse and shift are evolved according to the
``\emph{$1+\log$}''~\cite{Bona95b} and
``$\tilde{\Gamma}$-driver''~\cite{Alcubierre02a}, respectively,
including shift-advection terms which have proven to be important for
maintaining phase accuracy in the puncture motion~\cite{Baker:2005vv,
  vanMeter:2006vi}.  This combination of gauges are key ingredients of
the ``moving puncture'' method~\cite{Baker:2005vv, Campanelli:2005dd}
and the control of evolution of the coordinate singularity within the
black hole throat~\cite{Hannam:2006vv}.

The fields are discretize on regular locally Cartesian grids which
cover the neighborhood of the binary to some finite coordinate
radius. Finite differences at 8th-order are used to numerically
approximate derivatives. Boundaries between coordinate patches
are determined by 5th-order Lagrange polynomial interpolation.
Fields are evolved in time according to the BSSNOK prescription,
using a 4th-order Runge-Kutta integrator. A detailed description
of the numerical scheme can be found in~\cite{Pollney:2009yz}.

An objective of this paper is to compare the gravitational waveforms
measured at $\scri$ with the locally determined measures which are in
common use in numerical relativity. These local measurements generally
evaluate the geometrical fields (the metric and associated
derivatives) on a coordinate sphere of constant radius. The
measurement spheres are assumed to be in the ``wave-zone'', which is
commonly taken to be from $r=18M$~\cite{Gonzalez:2008} to
$r=225M$~\cite{Scheel:2008rj} from the source.  The particular
numerical grid structure employed in this paper follows that of
companion studies in which we have demonstrated accurate and
convergent wave measurements to $r=1000M$~\cite{Pollney:2009ut,
  Pollney:2009yz}. Polynomial extrapolations to spatial
$r\rightarrow\infty$ are applied in order to reduce the systematic
error associated with the finite radius measurements. Below, we
demonstrate that the error associated with this procedure in
comparison with the CCE result is, however, still larger than the
numerical error for the models we have studied.

Two methods are used in order to estimate the gravitational radiation
at a given radius. The first evaluates the Weyl tensor in a particular
null frame $(\mathbf{n}, \mathbf{l}, \mathbf{m},
\mathbf{\bar{m}})$~\cite{Penrose:1963}, which is chosen to be oriented
along the coordinate radial direction, and orthonormalised using the
local metric. According to the peeling property of asymptotically flat
spacetimes, the component
\begin{equation}
  \psi_4 := C_{\mathbf{n}\mathbf{\bar{m}}\mathbf{n}\mathbf{\bar{m}}}
\end{equation}
is associated with the outgoing gravitational radiation.
As an alternative to $\psi_4$, we also measure waves by evaluating
the Zerilli-Moncrief variables, assuming the spacetime metric at large
radius from the source to be representable as perturbations of a fixed
Schwarzschild background~\cite{Allen98a1,Rupright98,Camarda97c}. We
evaluate 1st-order gauge invariant odd-parity and even parity
multipoles, ($Q^\times_{lm}$ and $Q^+_{lm}$,
respectively)~\cite{Moncrief74, Abrahams95b}. The implementation of
the $\psi_4$ and Zerilli-Moncrief measurements follow the outline
described in~\cite{Pollney:2007ss, Pollney:2009yz}.

Our evolution code is built using the \texttt{Cactus} computational
framework \cite{Goodale02a, cactusweb1}.  The grid structures are
implemented using extensions to the \texttt{Carpet} mesh refinement
driver \cite{Schnetter-etal-03b, Schnetter06a, carpetweb}, allowing
the evolution domain to be covered by multiple patches with
generalized local coordinates~\cite{Pollney:2009yz}.
Initial data is computed using the \texttt{TwoPunctures}
code~\cite{Ansorg:2004ds}), which have been used
for previous Cartesian evolutions.

\section{Implementation of characteristic extraction as post-processing code}
\label{s-imp}

The characteristic extraction is independently performed as a post-processing step after the full Cauchy evolution.
During the Cauchy evolution, we output the Cauchy 4-metric and its
derivatives. The variables are decomposed in terms of spherical harmonic amplitudes on a set of extraction spheres with fixed radii around the
origin of the black hole spacetime.
On the characteristic side, this data can then be read and
reconstructed to generate inner boundary data on the world-tube.

Since we obtain the fields at $\scri$ in terms of coordinate time $u$, we finally have to transform the fields to constant Bondi time $u_B$.
Furthermore, due to the rather complicated procedure of calculating the conformal factor $\omega$ for the calculation of the news, we have implemented a linearized
approximation that turns out to be numerically more accurate than the full non-linear computation.

In the next subsections, we will describe these procedures in more detail.

\subsection{World-tube boundary data}

Boundary data for the characteristic evolution are constructed during Cauchy evolution in terms of the ADM variables on coordinate spheres with fixed radius $R$.
The time series of the coordinate spheres at successive times
generates the discretize world-tube.

In order to be able to transform the ADM variables to a representation of the full 4-metric in terms of Bondi coordinates and in terms of the characteristic evolution variables 
in a neighborhood of the world-tube, we have to compute the following quantities:
the ADM metric components $\gamma_{ij}$, as well as the time-derivatives $\p_t \gamma_{ij}$ and Cartesian derivatives $\p_k \gamma_{ij}$,
the lapse $\alpha$, as well as time and Cartesian derivatives $\p_t\alpha$ and $\p_i\alpha$, 
the shift vector $\beta^i$, and again time and Cartesian derivatives $\p_t\beta^i$ and $\p_i\beta^k$.
In the specific formulation of the evolution system that is used in this paper, we can compute these quantities as follows.
The BSSNOK evolution variables are transformed to the ADM variables after each evolution step and are therefore known everywhere.
What remains is the calculation of time and spatial derivatives.

The time derivatives of the lapse and shift vector are known directly
from the RHS of the ``1+log'' slicing condition and the hyperbolic
$\tilde{\Gamma}$-driver condition (see~\cite{Pollney:2009yz} for
details of the implementation of these gauge conditions). The RHSs of
these equations have to be computed at each time-step in order to
evolve the gauge variables with a method of lines (MoL) scheme, similar to the evolution
equations.

The time derivatives of the metric components can be obtained via the ADM relation
\be
\p_t \gamma_{ij}=-2\alpha K_{ij} + D_i\beta_j + D_j\beta_i\,,
\ee
where $D_i$ is the covariant derivative operator compatible with the spatial 3-metric $\gamma_{ij}$ and $K_{ij}$ the extrinsic curvature.
In order to compute the covariant derivative, it is necessary to invert the 3-metric and to calculate the Christoffel symbols on each grid point. Because the shift vector is given in contravariant form,
we have to transform it to its covariant representation with index
down. After looping over all points on the computational grid, we
store the outcome in grid functions, \ie,  one per metric component.

Next, obtain the Cartesian derivatives of the metric, lapse and shift.
This is straight-forwardly done by applying finite difference operators to the variables. Note that in case of non-Cartesian local coordinates, the derivatives are
calculated via the global derivatives as described in~\cite{Pollney:2009yz}.
For practical reasons (particularly storage and memory requirements), it is sufficient to calculate the radial instead of the Cartesian derivatives. The Cartesian derivatives can later be reconstructed 
in terms of angular derivatives of the spherical harmonics which are known analytically.
This saves two additional grid functions per ADM variable.
We obtain the radial derivatives in terms of the Cartesian derivatives via
\be
\p_r = \frac{1}{r} \left(x\p_x+y\p_y+z\p_z \right).
\ee

This completes the first step of calculating all variables that are necessary for the construction of characteristic boundary data.
What remains is the projection of these variables on to a coordinate sphere with fixed radius $R$ and a subsequent decomposition in terms of scalar spherical harmonics.
To project the variables on to the sphere, we use fourth-order Lagrange interpolation.
Afterwards, the variables are decomposed as
\be
f_{\ell m} = \int_{S^2} d\Omega\, \bar{Y}_{\ell m} f(\Omega)\,,
\ee
for all variables $f$ whether scalar, vector or tensor quantities.
The projection and decomposition are both done on polar spherical coordinate spheres where the surface integration is done with a Gauss quadrature scheme as described in \cite{Pollney:2009yz},
and the resulting array of spherical harmonic modes is stored in a file for further
processing.
The reconstruction from the data in that file will be described in the next subsection.

\subsection{Reconstruction from harmonic modes}

Once the boundary data in terms of the ADM variables and their time and radial derivatives are known on a coordinate sphere with radius $R$ and stored in a file, the characteristic
boundary module reads and interprets the data in that file in order to
reconstruct the variables on $S^2$. That is, the characteristic code is run as a post-processing tool
to obtain the gravitational radiation at $\scri$.
The variables are reconstructed via
\be
f = \sum_{\ell, m} f_{\ell m} Y_{\ell m}\,.
\ee
for any variable $f$.

The advantage of reconstructing the variables from harmonic modes is the independence in terms of angular coordinates,\footnote{On the Cauchy side, the ADM quantities are represented in polar-spherical coordinates
on $S^2$. The characteristic code, on the other hand, works with a stereographic coordinate mapping of $S^2$.}
as well as resolution.
Additionally, the necessary finite $\ell$-mode cut-off can act as
as a filter that factors out angular high-frequency noise. In
practice for the binary black hole models considered here, the
amplitudes of higher $\ell$ modes falls off rapidly so that they can
be considered essentially irrelevant numerically for, say, $\ell\ge
10$.

Note that for practical reasons, we have preferred to calculate the radial instead of the Cartesian derivatives of the ADM variables.
As a consequence, we have the harmonic modes of the radial derivatives. In order to obtain the Cartesian derivatives instead, one can
take angular derivatives of the spherical harmonics and then apply a Jacobian transformation from stereographic to Cartesian coordinates
\be
\frac{\p}{\p x^i} f = \frac{\p r}{\p x^i}\, \p_rf_{\ell m} Y_{\ell m}(q, p) 
   + f_{\ell m} \frac{\p q}{\p x^i} \frac{\p Y_{\ell m}}{\p q} + f_{\ell m}\frac{\p p}{\p x^i}\frac{\p Y_{\ell m}}{\p p}\,,
\ee
where $q,p$ denote stereographic angular coordinates.
By using the relation 
\be
A_{,q} = \frac{\eth A + \bar\eth A - 2ips A}{1+q^2+p^2}\,, \qquad A_{,p} = \frac{i(\bar\eth A - \eth A + 2qs A)}{1+q^2+p^2}\,,
\ee
for a quantity $A$ with spin-weight $s$,
we can express partial derivatives with respect to angular coordinates in terms of the eth-derivative
\bea
\frac{\p}{\p x^i} f &=& \frac{\p r}{\p x^i}\, \p_rf_{\ell m} Y_{\ell m} \nonumber \\
   &+& f_{\ell m} \frac{\p q}{\p x^i}  \left(\frac{\eth Y_{\ell m} + \bar\eth Y_{\ell m}}{1+q^2+p^2} \right)
     + f_{\ell m} \frac{\p p}{\p x^i} i\left(\frac{\bar\eth Y_{\ell m} - \eth Y_{\ell m}}{1+q^2+p^2}\right)\,,
\eea
where we have used the fact that the scalar spherical harmonics have spin-weight $s=0$.
This can be further simplified by replacing the eth-derivatives of the spherical harmonics by spin-weight $s=\pm1$ spherical harmonics.
Finally, we can write
\be
\frac{\p}{\p x^i} f = \frac{f^{\ell m}\sqrt{\ell(\ell+1)}}{1+q^2+p^2}\left[{}_1Y_{\ell m}(q_{,i}-ip_{,i})- {}_{-1}Y_{\ell m}(q_{,i}+ip_{,i}) \right] + r_{,i} f^{\ell m}_{,r}Y_{\ell m}\,,
\label{eq:radial-deriv}
\ee
where
\bea
r_{,i} &=& \p_i\sqrt{x^2+y^2+z^2} = \frac{x_i}{r}\,, \\
q_{,i} &=& \p_i\left(\frac{x}{\sqrt{x^2+y^2+z^2}\pm z} \right) = \frac{1}{(r\pm z)^2}\left(r\pm z-x^2/r,\, -xy/r,\, -xz/r\mp x \right),\\
p_{,i} &=& \p_i\left(\frac{\pm y}{\sqrt{x^2+y^2+z^2}\pm z} \right) = \frac{1}{(r\pm z)^2}\left(\mp xy,\, \pm r+z\mp y^2/r,\, -y\mp yz/r \right),
\eea
where the upper sign is valid for the north patch and the lower sign is valid for the south patch.
This means that we do not have to calculate any extra derivatives numerically and by virtue of Eq.~(\ref{eq:radial-deriv}), we immediately obtain the Cartesian derivatives from the radial derivatives.

Once the ADM variables and their derivatives are reconstructed on the coordinate sphere, the code executes the computations discussed in Sec.~\ref{sec:char-extr} to
obtain the boundary data in terms of the characteristic evolution variables and in terms of the Bondi coordinates.
Once this step is complete, the code can use the boundary data to evolve the characteristic evolution variables further in time.

Results for the application of this method to a binary black hole merger are presented in Sec.~\ref{sec:results}. However, the fields at $\scri$ first have to be transformed to constant Bondi time $u_B$.
This is described in the next subsection.

\subsection{Interpolation to constant Bondi time and mode decomposition}

After the evolution of the characteristic variables to $\scri$, the news function $N$ and the Weyl scalar $\Psi_4$ are computed in terms of inertial 
spatial Bondi coordinates $(r_B, p_B, q_B)$, but
are not yet given in terms of the inertial Bondi time coordinate $u_B$, \ie, we have $N=N(r_B, p_B, q_B, u)$. Also, one deals with the complete 2D data at $\scri$, but for analysis purposes, it
is more convenient to have the data decomposed in terms of spherical harmonic modes.

In a previous implementation, the transformation to constant $u_B$ was achieved by storing the news function as a function on $S^2$ at each time-step to a file.
In a post-processing step, these time-slices were then read and interpolated to constant Bondi time. This, however, is rather inconvenient to use in practice.
Rather, one would like to have the variables automatically transformed
to the correct time coordinate and decomposed in terms of spherical
harmonic modes, avoiding the need for additional post-processing tools.
For this reason, we have extended the module for calculating the news
function and the Weyl scalar $\Psi_4$ to take care of these issues.

During characteristic evolution, we know the inertial Bondi time in terms of coordinate time and angular coordinates $u_B=u_B(u, p, q)$ at each time-step at $\scri$.
We have found that in practice the difference between $u$ and $u_B$ is
very small, so that
by keeping five time levels of $u_B$, $N$ and $\Psi_4$ in memory at
each time-step, it is possible to interpolate $N$ and $\Psi_4$ to $u_B=\textrm{const.}$ for each point on $S^2$.
This is done  by means of fourth-order Lagrange interpolation on the
fly during evolution.
In practice, we average the Bondi time over $S^2$ on the past-past time-level to find the target time $u_B=\textrm{const}$. We use the past-past time-level to have the interpolation stencil centered around
the target interpolation time to maintain maximal accuracy in the interpolation.
Once the target time is known, $N$ and $\Psi_4$ are interpolated for each point on $S^2$ to $u_B=\textrm{const.}$ so that we finally have $N=N(r_B, p_B, q_B, u_B)$ and $\Psi_4=\Psi_4(r_B, p_B, q_B, u_B)$.

Afterwards and if requested by the user, the code can automatically decompose the two functions in terms of spin-weighted spherical harmonic modes.
Therefore, the data is available in a format that is convenient for analysis without any further post-processing. 
A set of post-processing tools
can read in the harmonic modes of $\Psi_4$ and calculate radiated energy, linear and angular momentum as well as the wave-strain $h$.

However, there is one word of caution required. As mentioned earlier, the
characteristic code computes a quantity $\Psi_4$ which is related to
the commonly used $\psi_4$ via Eq. (\ref{eq:psiPsi}).
It follows that the harmonic modes are related by
\be
\psi_4^{\ell m} = - 2\bar{\Psi}_4^{\ell,-m} (-1)^{m}\,.
\ee

\subsection{The linearized conformal factor}

In order to compute the news $N$ and $\Psi_4$, it is first necessary
to calculate the conformal factor $\omega$ of the conformal
compactification of the Bondi metric at $\scri$.  However, the full
non-linear computation of this factor can display a high level of
noise when it is implemented numerically, reducing its accuracy.
Since in all cases that we consider the fields at $\scri$ are in a
linear regime, we can make use of a relation between the conformal
factor $\omega$ and $J$ at $\scri$ in the linearized approximation.

Given $J$ at $\scri$ in terms of $s=2$ spin-weighted spherical harmonics, \ie,
\be
J\vert_{\scri}(u, x^A)=\sum_{\ell \geq 2,m} J_{\ell,m}(u){}_2 Y_{\ell,m}\,,
\ee
we can write $\omega$ in terms of scalar spherical harmonics as
\be
\omega(u,x^A)=1+\sum_{\ell \geq 2,m} \omega_{\ell,m}(u)Y_{\ell,m}\,,
\ee
where
\be
\omega_{\ell,m}(u)=-\frac{1}{4}\sqrt{\frac{\ell(\ell+1)}{(\ell-1)(\ell+2)}}\left(J_{\ell,m}(u) + (-1)^m \bar{J}_{\ell,-m}(u) \right)\,.
\ee

The linearized computation has been implemented as an alternative to
the full non-linear computation of $\omega$ and either method can be
selected at run-time.

\section{Tests of the characteristic extraction code}
\label{s-test}

In this section, we report on tests we have performed in order to calibrate the characteristic extraction code.
For this, we put an analytical solution to the ADM metric variables on the Cauchy side, generate the associated spherical harmonic modes
on each timestep and afterward run the characteristic extraction code.

\subsection{Linearized solutions}

We have tested convergence of the entire extraction process against an analytical (linearized) solution \cite{Bishop-2005b} by
transforming this solution to the ADM variables, \ie, lapse $\alpha$, shift $\beta^i$ and the 3-metric components $\gamma_{ij}$, on the
Cauchy side. According to the description in Sec.~\ref{s-imp}, we decompose the ADM variables on a world-tube after each timestep in order to produce
spherical harmonic modes that can be used as boundary data for the subsequent characteristic evolution.
On the characteristic side, we measure the error, that is, the difference between evolved variables and exact fields.
We do so by performing a set of two resolutions, and we can assume convergence against the linearized solution given that the 
amplitudes of the fields are small, \ie, they are in the linearized regime.

The solutions to be used are those given in
Sec.~4.3 of ref.~\cite{Bishop-2005b} for the case of a dynamic spacetime on
a Minkowski background with $\ell=2,\, m=0$. We write
\begin{eqnarray}
J&=& \sqrt{(\ell -1)\ell(\ell+1)(\ell+2)}\;{}_2Y_{\ell m} \Re(J_\ell(r) e^{i\nu u}), \nonumber \\
U&=& \sqrt{\ell(\ell+1)}\;{}_1Y_{\ell m} \Re(U_\ell(r) e^{i\nu u}),
\nonumber \\
\beta&=& Y_{\ell m} \Re(\beta_\ell e^{i\nu u}),  \nonumber \\
\; W_c&=& Y_{\ell m} \Re(W_{c\ell}(r) e^{i\nu u}),
\label{e-an}
\end{eqnarray}
where $J_\ell(r)$, $U_\ell(r)$, $\beta_\ell$, $W_{c\ell}(r)$ are in general
complex, and taking the real part leads to $\cos(\nu u)$ and $\sin(\nu u)$
terms. The quantities $\beta$ and $W_c$ are real; while $J$ and $U$ are
complex due to the terms ${}_2Y_{\ell m}$ and  ${}_1Y_{\ell m}$,
representing different terms in the angular part of the metric. 
In the case $\ell=2$
\begin{eqnarray}
\beta_2&=&\beta_0, \nonumber \\
J_2(r)&=&\frac{24\beta_0 +3 i \nu C_1 - i \nu^3 C_2}{36}+\frac{C_1}{4 r}
       -\frac{C_2}{12 r^3}, \nonumber \\
U_2(r)&=&\frac{-24i\nu \beta_0 +3 \nu^2 C_1 - \nu^4 C_2}{36} +\frac{2\beta_0}{r}
       +\frac{C_1}{2 r^2} +\frac{i\nu C_2}{3 r^3} +\frac{C_2}{4 r^4}, \nonumber \\
W_{c2}(r)&=&\frac{24i\nu \beta_0 -3 \nu^2 C_1 + \nu^4 C_2}{6} 
           +\frac{3i\nu C_1 -6\beta_0-i\nu^3 C_2}{3r}
           -\frac{\nu^2 C_2}{r^2} +\frac{i\nu C_2}{r^3} +\frac{C_2}{2r^4},
\label{e-NBl2}
\end{eqnarray}
with the (complex) constants $\beta_0$, $C_1$ and $C_2$ freely specifiable.


We present results for the case $\ell=2,\, m=0,\, \nu=1$, with
\begin{eqnarray}
C_1=10^{-6}, \qquad C_2=3\times10^{-6}, \qquad \beta_0=0
\end{eqnarray}
in Eq.~(\ref{e-NBl2}).

The numerical simulations use a compactified radial coordinate, Eq. (\ref{eq:compact}), with $r_{\rm wt}=9$. Data is prescribed at time $u=0$
and extracted at a number of different world-tube locations, specifically at $r=\{30,60,90,120\}$.
The stereographic grids (with ghost zones excluded)
are
\begin{equation}
\mbox{Coarse: } n_x=n_q=n_p=41, \qquad \mbox{ Fine: } n_x=n_q=n_p=81;
\end{equation}
and there is no overlap between the two patches, \ie, we set the
code parameter $q_{\mbox{size}}=1$ which means that on the nominal
grid, the holomorphic coordinate function $\zeta=q+ip$ takes values
in $q,p\in [-1,1]$. The compactified coordinate takes values $x\in[x_{\rm in},1]$ such
that the world-tube is located on the characteristic grid and close to the inner boundary to avoid
accuracy losses. For example, for a world-tube located at $r=30M$, we set $x_{\rm in}=0.74$. 
For different world-tube locations, $x_{\rm in}$ may be changed accordingly.

The grids on the Cauchy side have a radial resolution of $h=0.4M$ and $h=0.2M$
with the number of angular points per direction and per patch set to $n=21$ and $n=41$, respectively.
The timestep size is taken to be $\Delta t = 0.1M$ and $\Delta t = 0.05M$.

\begin{figure}
  \centering
  \includegraphics[width=8.0cm]{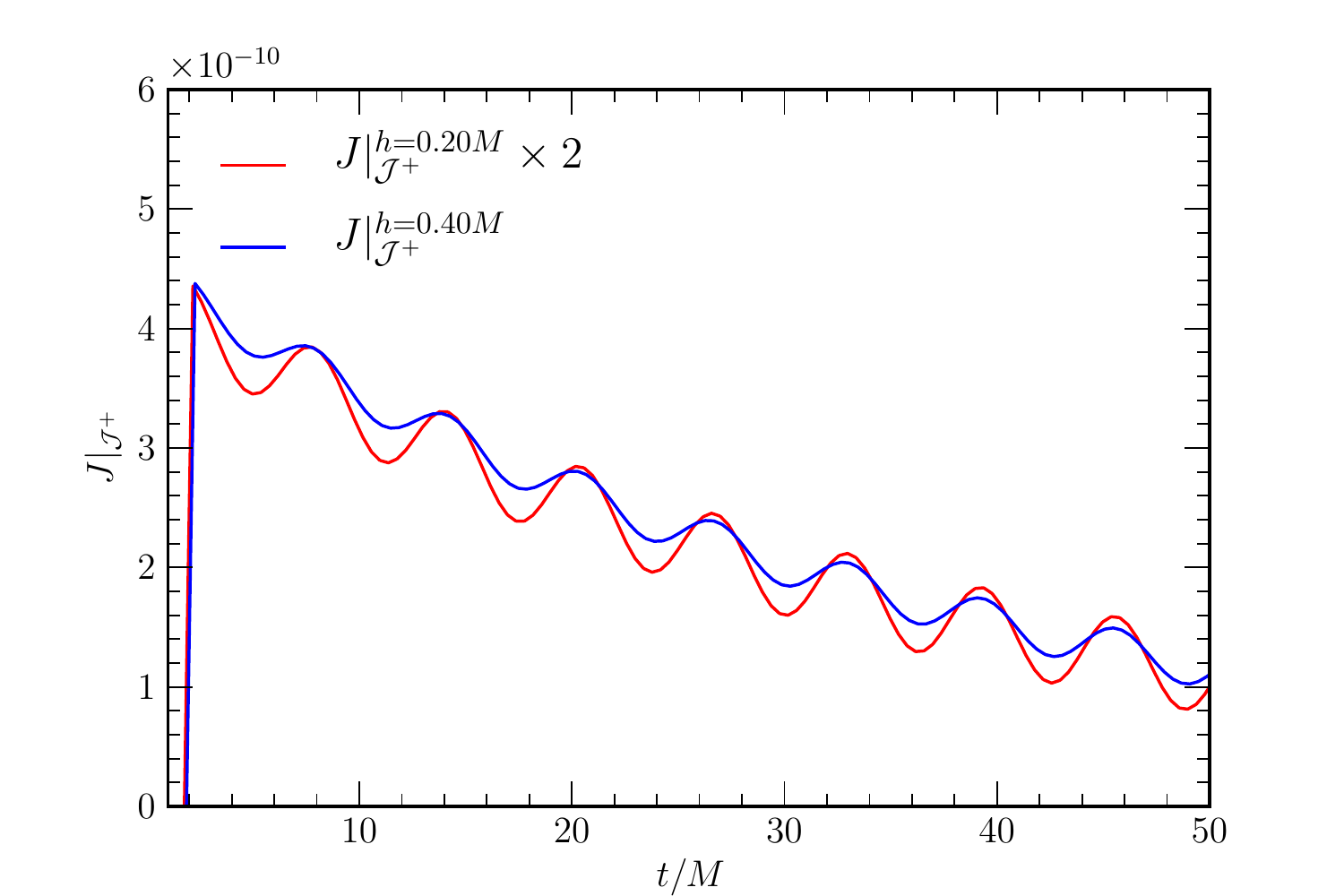}\hfill
  \caption{First-order convergence in the $L_2$-norm of $J$ at $\scri$ against a linearized solution as given by (\ref{e-NBl2}).}
    \label{fig:lin-conv}
\end{figure}

In Fig.~\ref{fig:lin-conv}, we show convergence in the $L_2$-norm of $J$ at $\scri$ to the linearized solution, starting off from a
world-tube at $r=30M$. We observe clear first-order convergence. This
is consistent with the fact that the boundary data for $U_{,r}$ is only
known to first-order off the world-tube. However, for other models
presented in subsequent sections, it appears that the coefficient of
numerical error arising from $U_{,r}$ is small enough that a
second-order convergence exponent can be measured, corresponding to
the interior finite difference accuracy of the characteristic code.

\subsection{Gauge invariance of a shifted Schwarzschild black hole}

\begin{figure}
  \centering
  \includegraphics[width=8.0cm,trim=0 0 0 0,clip=true]{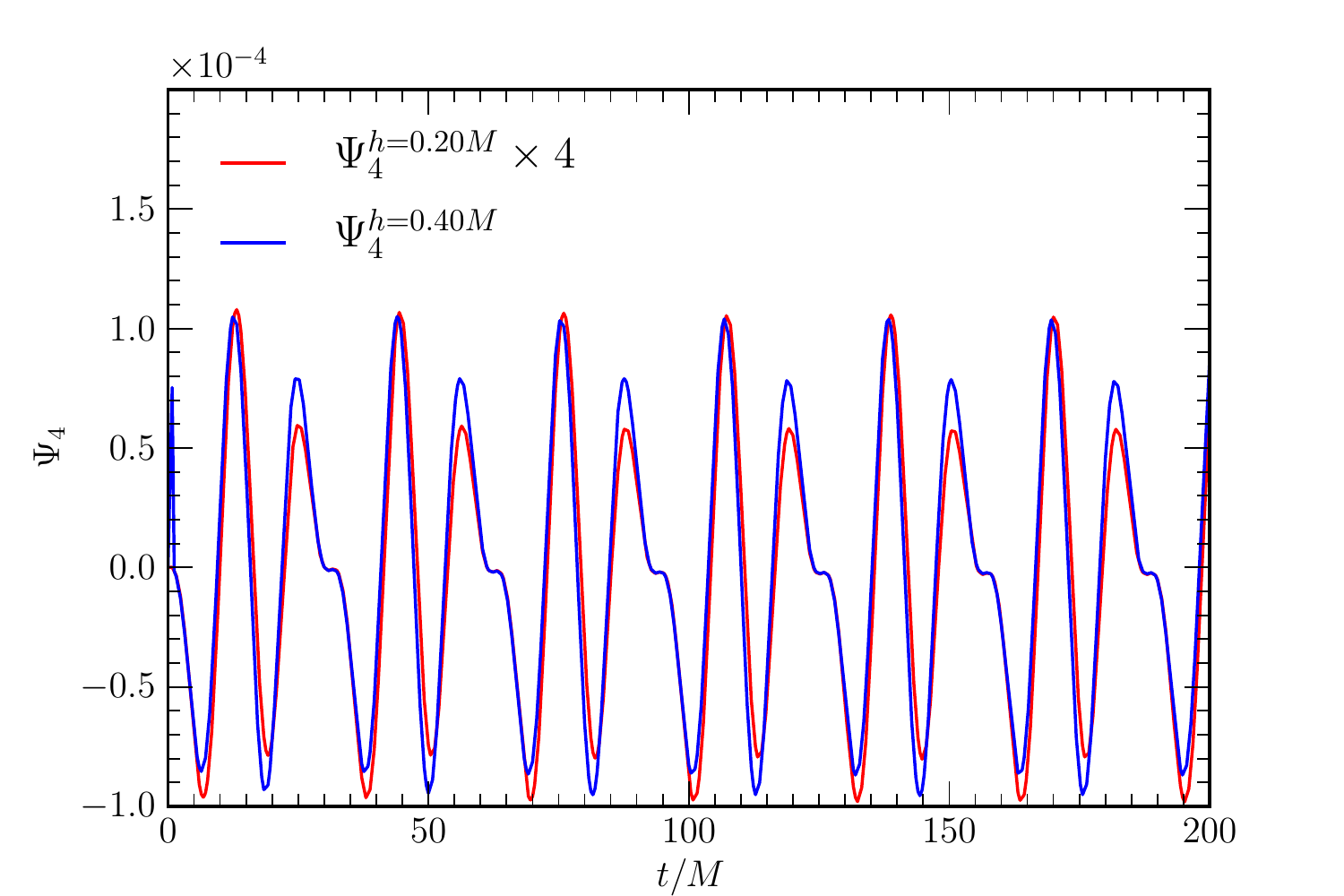}\hfill
  \includegraphics[width=8.0cm,trim=0 0 0 0,clip=true]{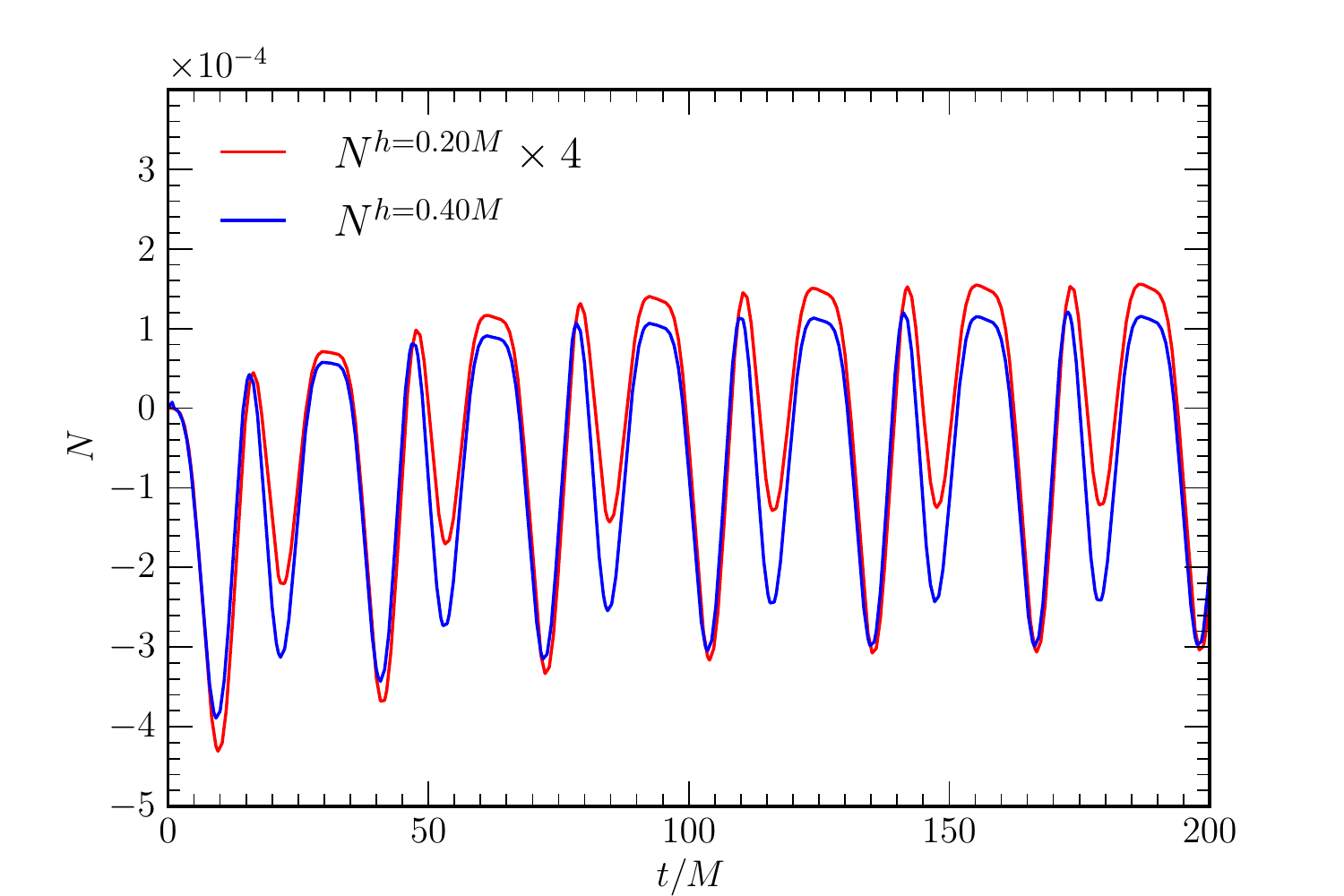}\hfill
  \caption{Second-order convergence of the gravitational wave signal to zero for the
    shifted-Schwarzschild test case. The \textit{left panel} shows the $\ell=2, m=0$ mode of 
           the Weyl scalar $\Psi_4$ for the two resolutions $h=0.40M$ and $h=0.20M$ where the latter is scaled for second-order convergence.
           The \textit{right panel} shows the $(\ell,m)=(2,0)$ mode of the news $N$ for the same resolutions and the same rescaling. Similar plots for all other modes exhibit the same order of convergence.}
    \label{fig:os-conv}
\end{figure}

An important aspect of CCE as a wave extraction method is the fact
that it should produce invariant results independent of the gauge
at the world tube.
As a simple test problem which demonstrates this property, we place a single static Schwarzschild black hole at the origin of our spacetime, and
apply a shift to the solution, according to
\be
\beta^x =  A\omega\cos(\omega t)\,, \qquad \beta^y=\beta^z=0\,,
\ee
so that the x-coordinate oscillates as
\be
x = A\sin(\omega t)\,,
\ee
where the amplitude is set to $A=1.0$ and the frequency is set to $\omega=0.1$.
As the spacetime is spherically symmetric and static, the resulting radiation content must be zero, 
and hence, in numerical simulations the residual in $\Psi_4$ and the news $N$ must converge to zero.
This is indeed what we find. Fig.~\ref{fig:os-conv} shows the $(\ell,m)=(2,0)$ mode of $\Psi_4$ and $N$ at two resolutions $h=0.4M$ and $h=0.2M$ scaled for
second-order convergence. All other modes converge to zero at the same order.

\begin{figure}
  \centering
  \includegraphics[width=8.0cm]{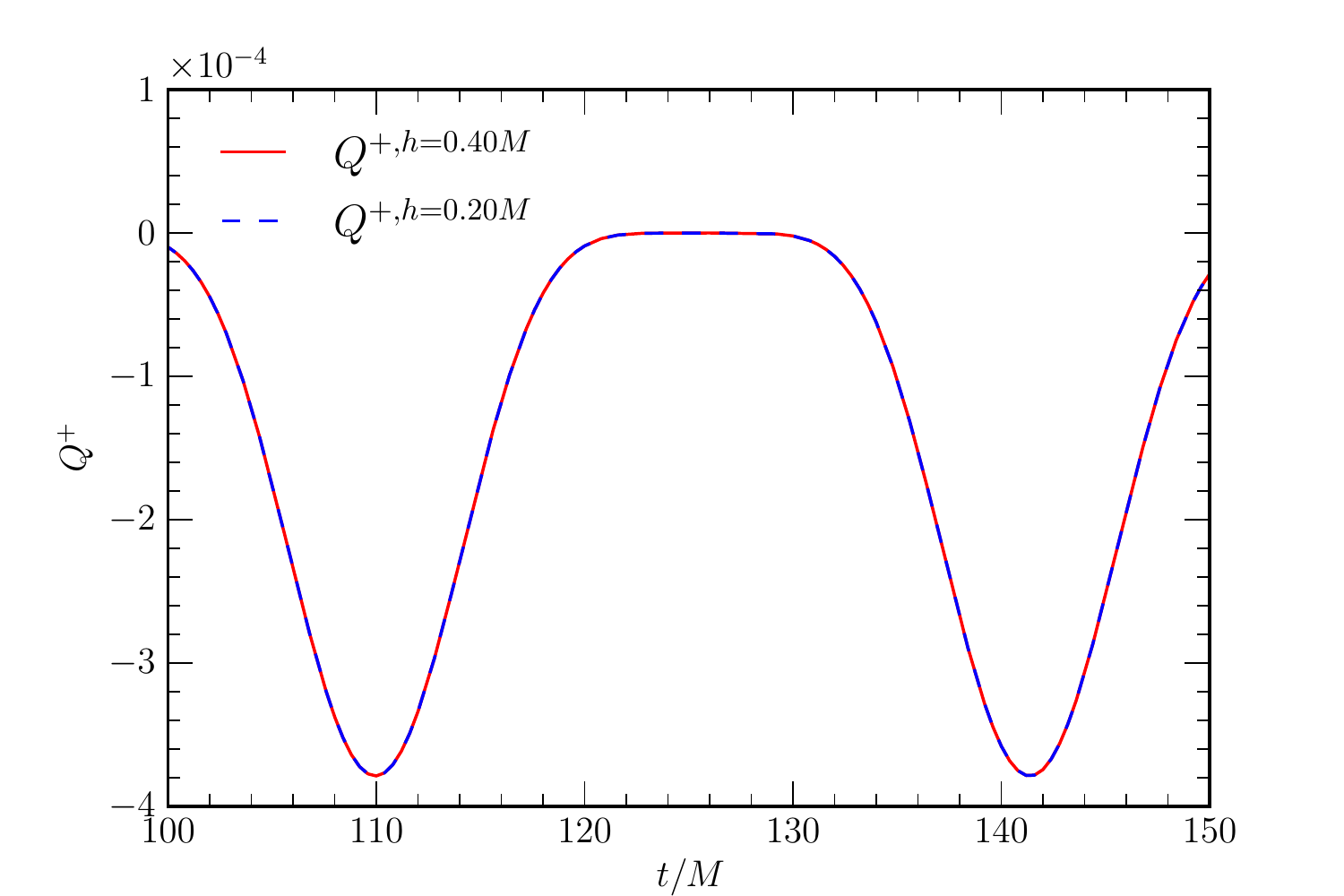}\hfill
  \includegraphics[width=8.0cm]{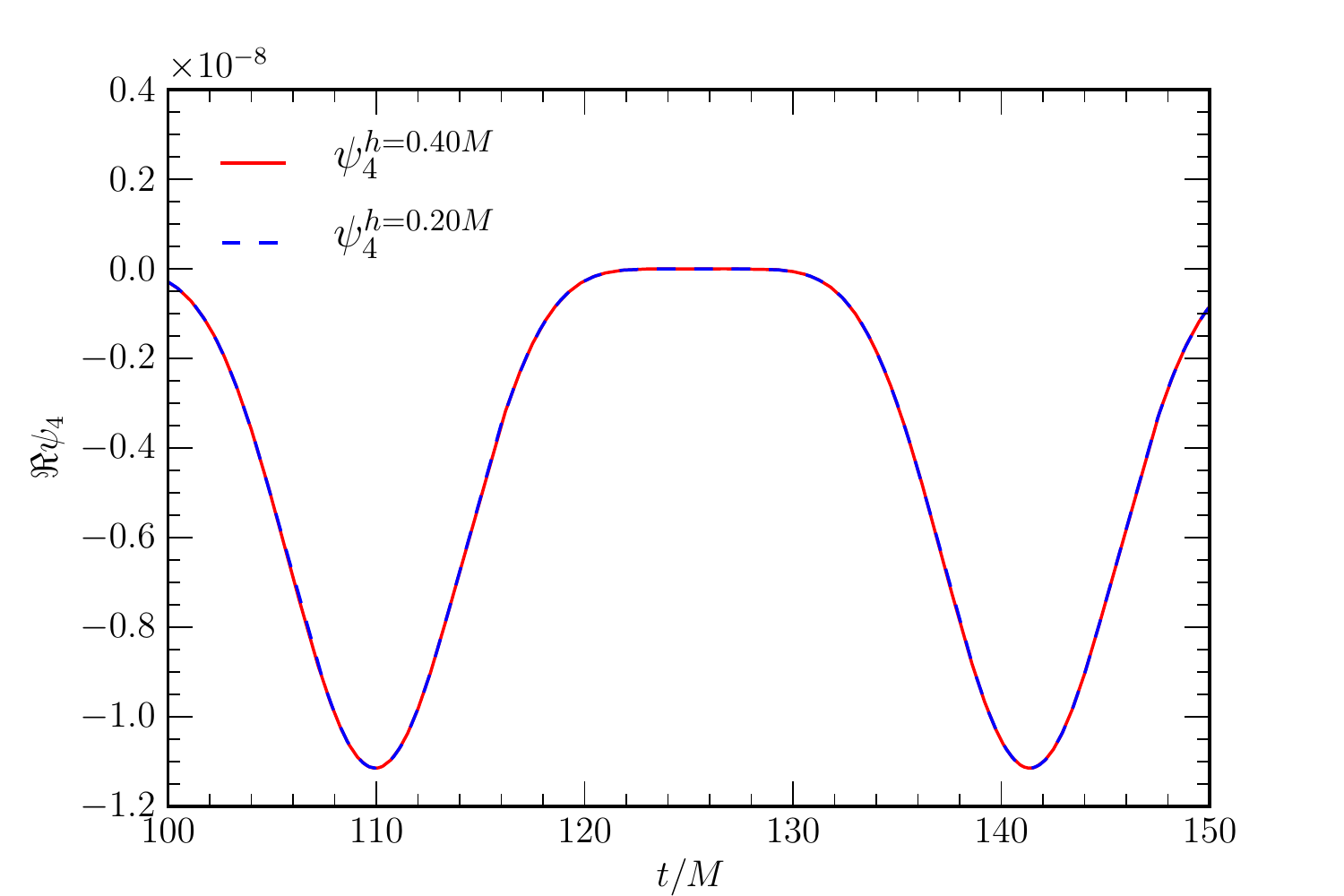}\hfill
  \caption{Non-convergent behavior of finite-radius measures of the
   gravitational radiation for the shifted-Schwarzschild test
   case. Shown are the perturbative Zerilli-Moncrief (even) master function $Q^+$ (\textit{left}) 
           and $\Re\psi_4$ computed at a finite radius (\textit{right}) extracted at $r=100M$. 
           For each case, we plot two resolutions $h=0.20M$ and $h=0.40M$ of the
           $(\ell,m)=(2,0)$ mode without any rescaling. We find that
           the waveforms do not converge to zero, as should be expected
           for this test case.}
    \label{fig:os-non-conv-Q}
\end{figure}

In contrast to this, finite radius extraction methods based on $\psi_4$ and on gauge-invariant perturbations of Schwarzschild fail to be convergent.
This is shown in Fig.~\ref{fig:os-non-conv-Q}, where we plot
the perturbative Zerilli-Moncrief (even) master function $Q^+_{\ell=2, m=0}$ \cite{Moncrief74, Zerilli70,
Nagar:2005ea} and $\psi_4$ for the two resolutions $h=0.4M$ and $h=0.2M$
without any rescaling.
The two waves are on top of each other indicating that the results do
not converge to zero.

Although the effect is exaggerated by this artificial test, the point
remains that the commonly used finite radius extraction methods are
susceptible to gauge variations which are difficult to disentangle
from the genuine wave signal. The CCE method as implemented removes
this systematic error completely.

\section{Results for binary black hole inspirals}
\label{sec:results}

\begin{figure}
  \begin{center}
    \includegraphics[width=86mm,clip,trim=0 0 0 0]{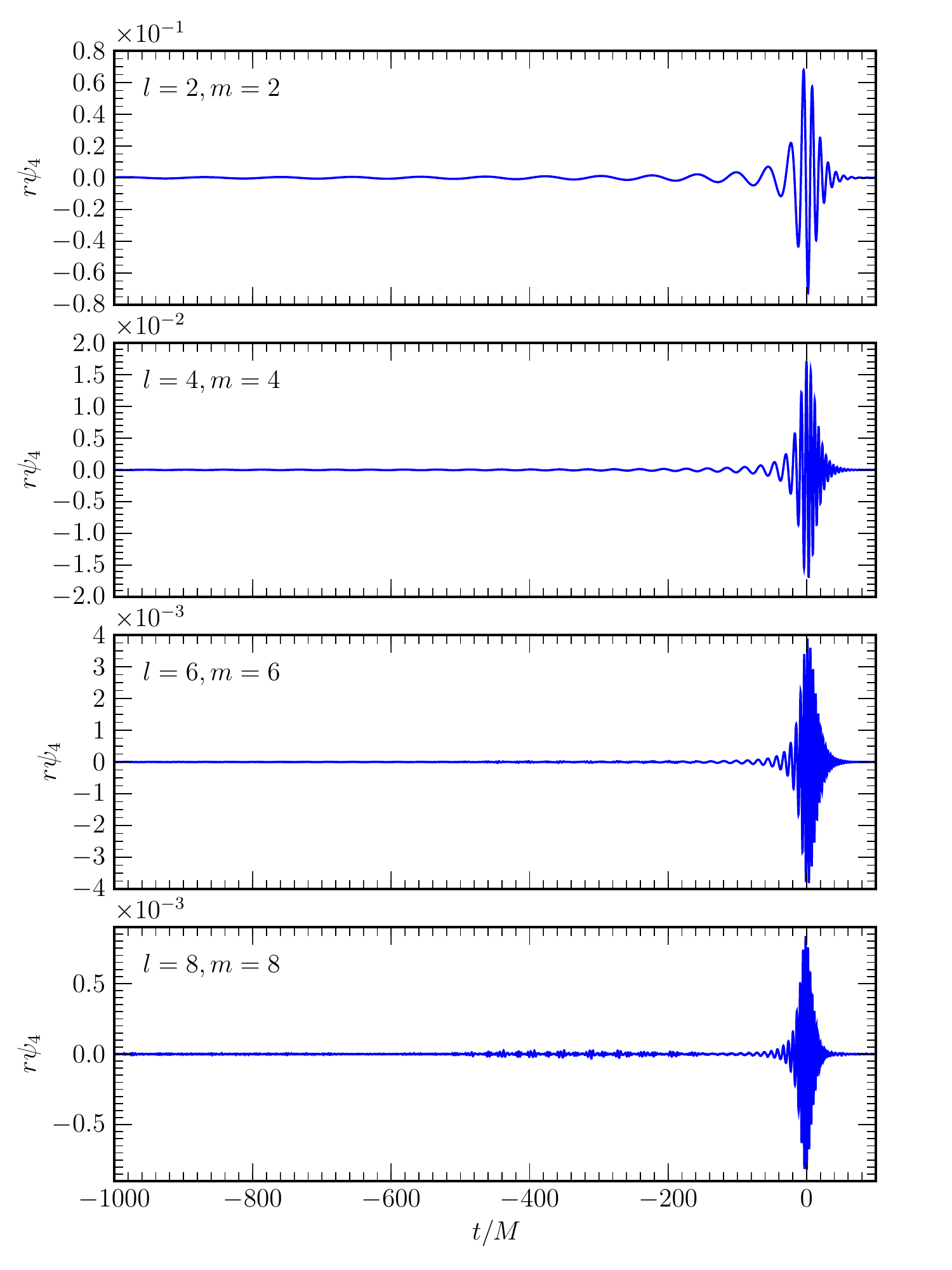}
  \end{center}
  \caption{Dominant spherical harmonic modes of $\psi_4$ for $\ell\le 8$  extracted at $\scri$ for the model $a_1=a_2=0$.}
  \label{fig:psi4-modes}
\end{figure}

We have carried out full three-dimensional, non-linear evolutions of two binary black hole systems for the whole spacetime
out to $\scri$ using the Cauchy and characteristic formulations as described in the previous sections.
The first configuration represents an equal-mass non-spinning binary, with modes
up to $\ell=8$ of the CCE waveform illustrated in Fig.~\ref{fig:psi4-modes};
while the second configuration is an equal-mass binary
with individual spins that are aligned with the orbital angular momentum and hence the system is not subject to
precession effects.
In this section, we compare the differences in the waveform measures at finite radius and at $\scri$, and compute various 
features of the two binaries.

We demonstrate that for these tests, the wave-extraction at $\scri$ contains only numerical error. 
Additionally, we measure the energy, linear and angular momenta radiated through gravitational waves, both at finite radii and at $\scri$
and test for conservation of energy and angular momentum.
As the emission of gravitational radiation is symmetric in the first case, but anti-symmetric in the second case, the net-linear
momentum carried by the gravitational waves is non-zero for the second binary.
Hence, this binary gives rise to a gravitational recoil, or kick, of the merger remnant.
We measure this effect to high precision using gravitational waveform data at finite radii and at $\scri$.

\subsection{Initial data}

For the first equal-mass non-spinning binary, the numerical evolution starts from an initial separation $d/M=11.0$,
and goes through approximately 8 orbits (a physical time of around
$1360M$), merger and ring-down. The masses of the punctures are set to
$m=0.4872$ and are initially placed on the $x$-axis with momenta
$p=(\mp 0.000728, \pm0.0903, 0)$, giving the initial slice an ADM mass
$M_{\rm ADM}=0.99051968 \pm 2\times 10^{-8}$. 
The z-axis is chosen to coincide with the initial orbital angular momentum and is given in Table~\ref{tab:binaries}.
The y-axis is then chosen perpendicular to the other axis.

The second binary has been investigated mainly so as to be able to measure the kick-velocity that the merger remnant
acquires due to the asymmetric emission of gravitational radiation.
As this effect depends largely on the very last orbit (compare \eg \cite{Pollney:2007ss}), it is sufficient
to consider a binary with a closer initial separation which in our case has been chosen to be $d/M=4.0$.
The initial masses of the punctures are set to $m=0.2999$ and are initially placed on the x-axis with momenta
$p=(\mp 0.00211428, \pm 0.112539, 0)$, giving the initial slice an ADM mass
$M_{\rm ADM}=0.98826246 \pm 2\times 10^{-8}$. 

The coordinate system is chosen such that the initial orbital angular momentum is aligned with the z-axis of the coordinate system, and we
represent the initial spin vectors by
\be
\mathbf{S}_1=(0,0,a_1)\,, \qquad \mathbf{S}_2=(0,0,a_2)\,,
\ee
where we have chosen $(a_1,a_2)=(0,0)$ for the first binary and $(a_1,a_2)=(0.8,-0.8)$ for the second one.

The initial data parameters were determined
using a post-Newtonian evolution from large initial separation,
following the procedure outlined in \cite{Husa:2007rh}, with the conservative
part of the Hamiltonian accurate to 3PN, and radiation-reaction to
3.5PN. The measured eccentricity of the resulting orbits is
$0.004\pm0.0005$

The most convenient way to set the initial data on the initial
null-hypersurface of the characteristic grid is to set $J(u=0)=0$ everywhere,
since the wave-extraction algorithm requires $J(u=0)_{\vert\scri}=0$.
Fortunately, it turns out that $J=0$ on the extraction world-tube $\Gamma$
at $u=0$ is consistent with conformal flatness, although in general there
will be an inconsistency at points on the null cone off the initial Cauchy
hypersurface.

\subsection{Grid setup}

The binary black hole evolution was carried out on a 7-patch grid
structure, as described in \cite{Pollney:2009yz, Pollney:2009ut},
incorporating a Cartesian mesh-refined region which covers the
near-zone, and six radially oriented patches covering the wave-zone.
The inner boundary of the radial grids was placed at $r_t=35.2M$
relative to the centre of the Cartesian grid.
\begin{figure}
  \begin{center}
    \includegraphics[width=120mm,clip,trim=0 0 0 0]{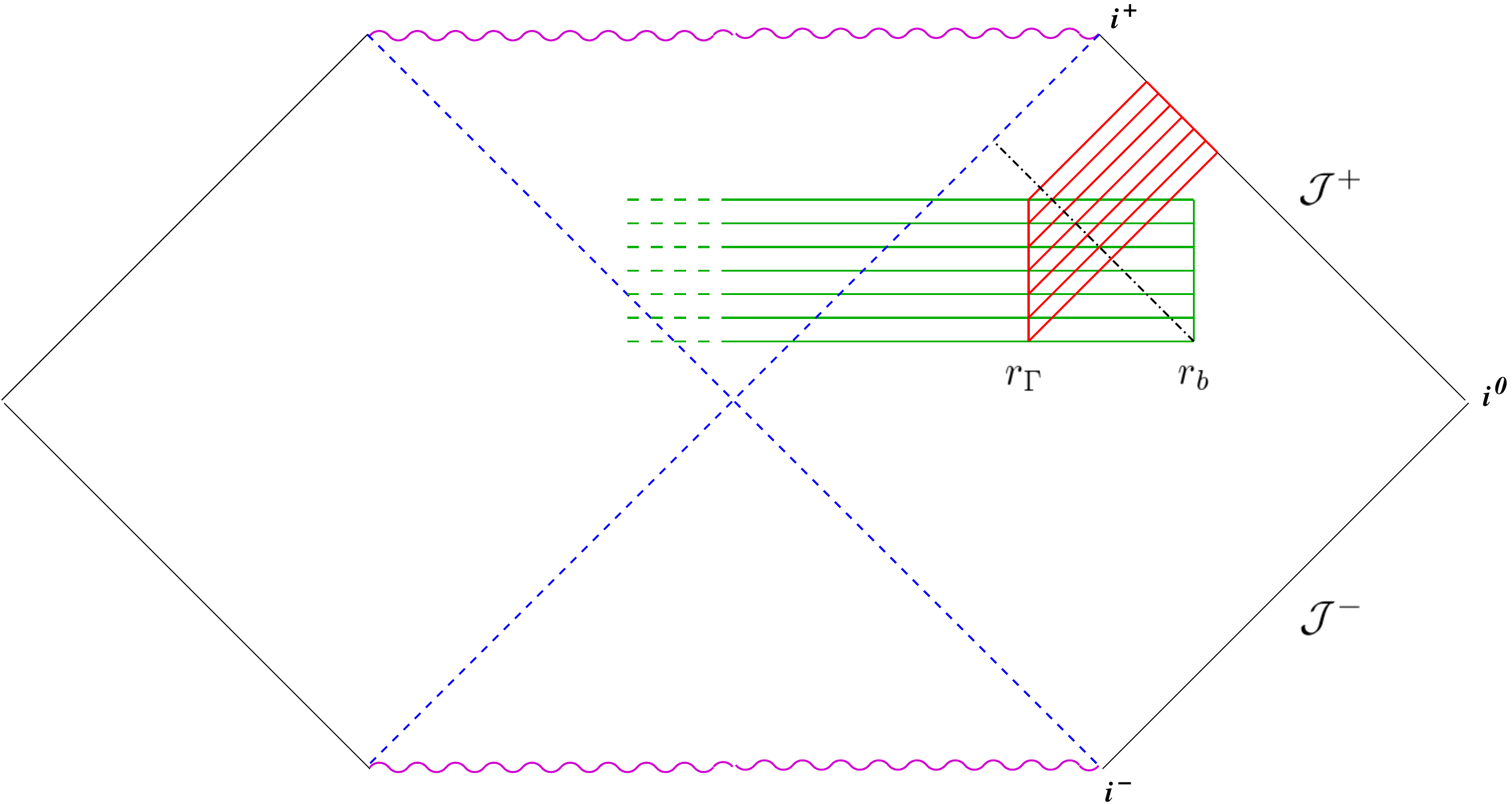}
  \end{center}
  \caption{Portions of the Kruskal diagram which are determined by the
    numerical evolution. The green horizontal lines indicate the region
           of spacetime that is determined by the Cauchy evolution and which has finite spatial extent with artificial outer boundary at $r_b$.
           The red diagonal lines indicate the region that is
           determined by characteristic evolution, and which has an
           inner boundary at 
           a world-tube $\Gamma$ located at $r_\Gamma$, consisting of data from the intersecting slices (green horizontal lines) of the Cauchy evolution.
           Note that due to the final spatial extent of the latter, the future Cauchy horizon of the Cauchy initial data, indicated by the dotted diagonal line parallel to $\scri$,
           principally prohibits the determination of the spacetime out to timelike infinity $i^+$. However, as long as the world-tube $\Gamma$ is located within the future Cauchy horizon, 
           the numerically evolved subset of the spacetime is determined consistently. Particularly, the artificial outer boundary $r_b$ is sufficiently removed such that the 
           evolved binary black hole spacetime forms a merged black hole before the Cauchy horizon reaches the world-tube.}
   \label{fig:causal_bc}
\end{figure}
The outer boundary for the spherical grids was chosen based on the
expected time duration of the measurement and radius of the furthest
extraction world-tube, in order to remove any influence of the artificial outer
boundary condition. In particular, given that the evolution takes a
time $T_m$ for the entire inspiral, merger and ringdown, and
a world-tube location at a finite radius $r_\Gamma$, we
would like to ensure that a disturbance traveling at the speed of
light $c=1$ \cite{Brown2007a,
  Brown2007b} from the outer boundary does not reach the world-tube radius
(see Fig.~\ref{fig:causal_bc})\footnote{The
$1+\log$ slicing condition which we use propagates at
$\sqrt{2}c$~\cite{Bona95b}, however this is a gauge mode and empirically we
find it to have negligible effect on finite radius measurements. Particularly, since CCE is gauge invariant, this does not affect the waveforms at $\scri$.}. Thus we place the
boundary at
\begin{equation}
  r_b > T_m + 2 r_\Gamma.
\end{equation}
For the particular evolutions considered here, for the non-spinning binary
and for the spinning binary respectively, we have evolution times
$T_m\simeq 1350M$ and $T_m\simeq 550M$, and
outermost extraction world-tubes $r_\Gamma=1000M$ and $r_\Gamma=500M$. We have placed
the outer boundary of the evolution domains at $r_b=3600M$ and $r_b=2000M$,
respectively.

The near-zone grids incorporate 5 levels of 2:1 mesh refinement,
covering regions centred around each of the black holes. For the
highest resolution we have considered here, the finest grid (covering
the BH horizon) has a grid spacing of $dx=0.02M$.  The wave-zone grids
have an inner radial resolution which is commensurate with the coarse
Cartesian grid resolution, $dr=0.64M$ in this case. This resolution is
maintained essentially constant to the outermost measurement radius
($r=1000M$), at which point we apply a gradual decrease in resolution
(see \cite{Pollney:2009yz}) over a distance of $r=500M$.
From $r=1500M$ to the outer boundary, we maintain a resolution of
$dx=2.56M$, sufficient to resolve the inspiral frequencies of the
dominant $(\ell,m)=(2,2)$ mode of the gravitational wave signal.  The
angular coordinates have 31 points (30 cells) in $\rho$ and $\sigma$ on
each of the 6 patches. The time-step of the wave-zone grids is
$dt=0.144$, and we take wave measurements at each iteration.

We have carried out evolutions at three resolutions, $h=0.64M$, $h=0.80M$ and $h=0.96M$, in order to
estimate the convergence of our numerical methods (compare \cite{Pollney:2009yz}).

The characteristic extraction world-tubes $\Gamma$ have been placed at
$r_\Gamma=100M$ and $r_\Gamma=200M$ for the first binary, while for the second binary we have chosen $r_\Gamma=100M$ and $r_\Gamma=250M$.
We note that over the course of the evolution, these world-tubes are causally disconnected from the artificial outer boundary.

\subsection{Invariance with respect to world-tube location}

\begin{figure}
  \centering
  \includegraphics[width=8.0cm]{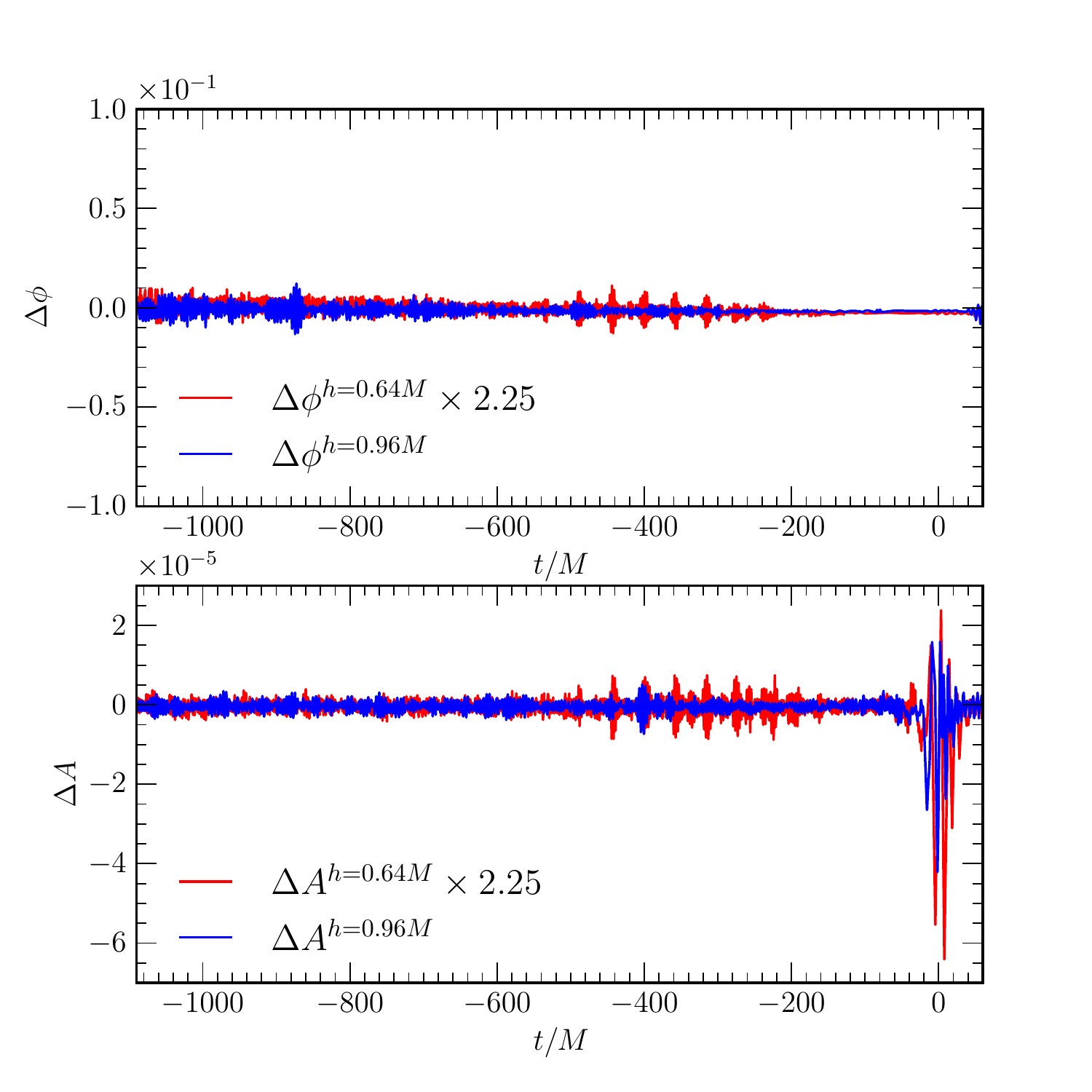}
  \includegraphics[width=8.0cm]{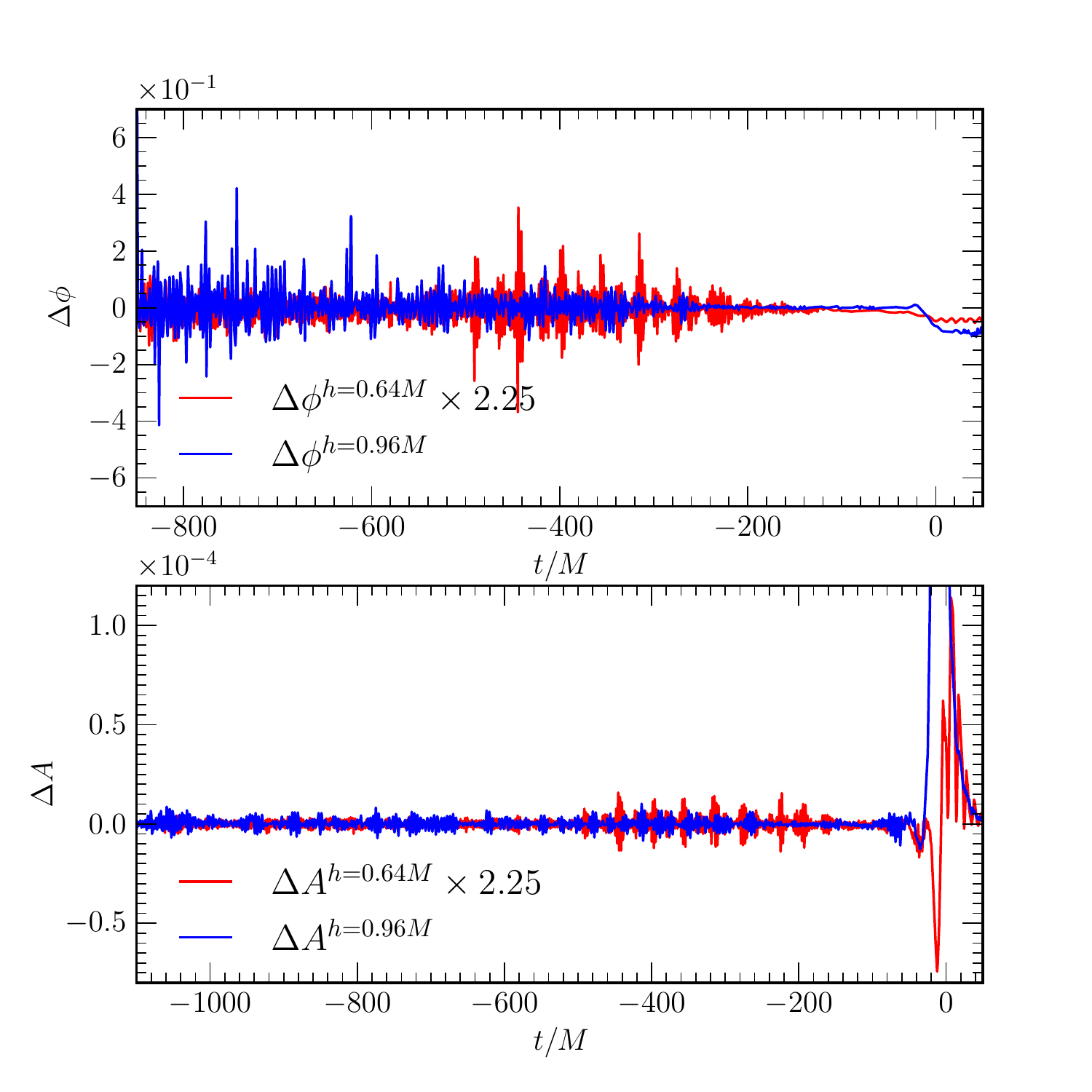}
  \includegraphics[width=8.0cm]{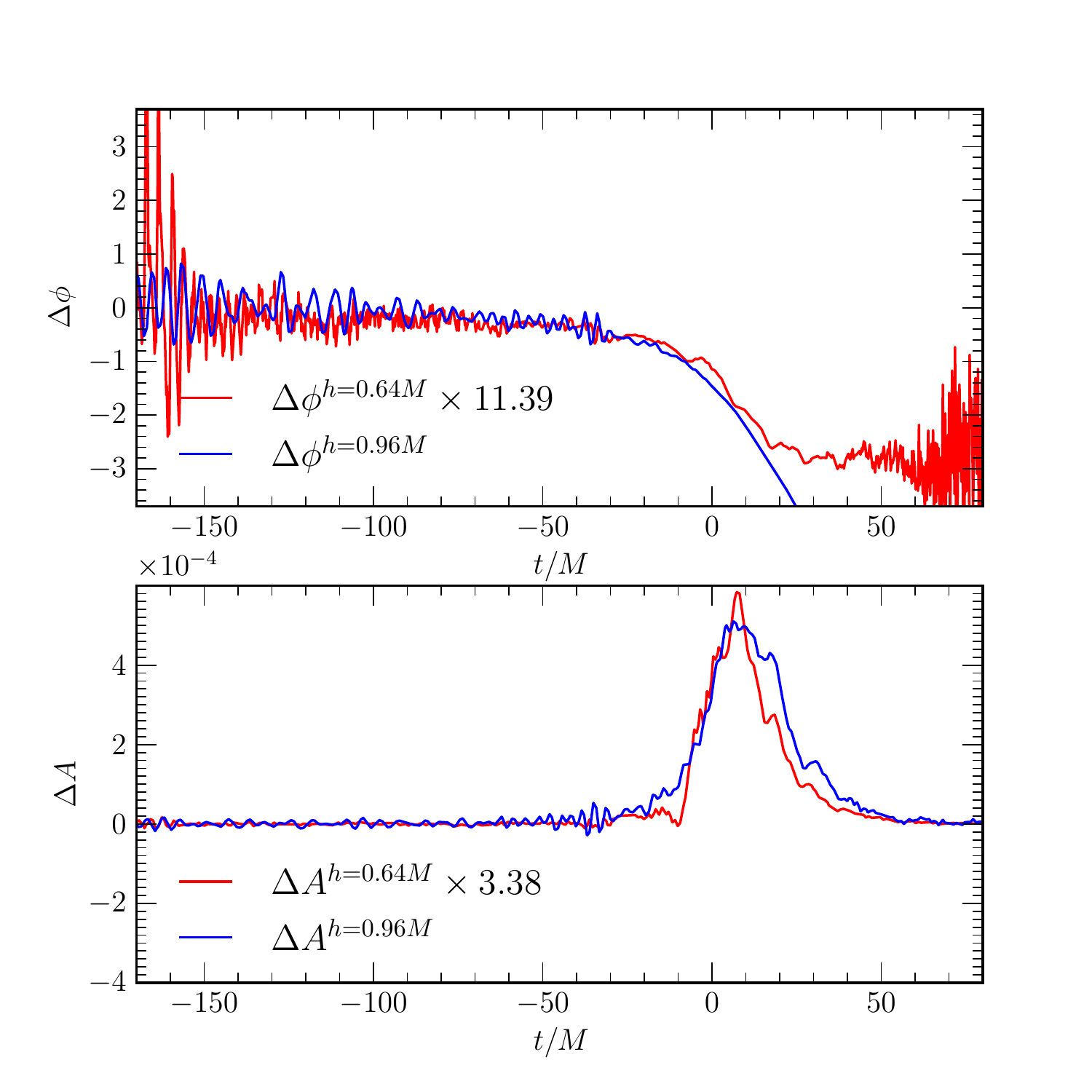}
  \caption{Convergence of the differences in wave-signal to zero for
    various test cases. The \textit{top left panels} show the amplitude and phase 
           differences between world-tube data at $r_{\Gamma}=100M$ and $r_{\Gamma}=200M$ of the $(\ell,m)=(2,2)$ mode of 
           the Weyl scalar $\psi_4$ for the two resolutions $h=0.96M$ and $h=0.64M$ where the latter is scaled for second-order convergence.
           The \textit{top right panels} show the the same for the $(\ell,m)=(4,4)$ mode of $\psi_4$.
           The \textit{lower panels} show convergence of the $(\ell,m)=(6,6)$ mode. The phase error for the high resolution is rescaled for sixth-order convergence
           while the amplitude error is rescaled for third-order convergence.}
    \label{fig:wave-conv}
\end{figure}

Invariance with respect to the world tube location is
demonstrated in Fig.~\ref{fig:wave-conv}. We have
considered the differences between waveforms at $\scri$ resulting from
two independent characteristic evolutions using boundary data at
$r_{\Gamma}=100M$ and $r_{\Gamma}=200M$, respectively, and for two
resolutions, $h=0.96M$ and $h=0.64M$. The difference between the
results should be entirely due to the discretisation error, and indeed
this is what we find. The differences converge to zero with approximately
second-order accuracy, as expected for the null evolution code. The figure
displays the differences in the amplitude and phase of the
wave mode $\psi_4(\ell=2,m=2)$ for resolution $h=0.96M$ and $h=0.64M$
scaled for second-order convergence. The same order of convergence is
also obtained for higher order modes such as $\psi_4(\ell=4,m=4)$ and $\psi_4(\ell=6,m=6)$ during inspiral.
However, the higher modes are more sensitive to the order of convergence on the Cauchy side during merger, and hence, the
error goes down with a higher convergence order (note that we use 8th-order finite difference operators).
The phase of $(\ell,m)=(6,6)$ mode converges at sixth-order while the amplitude converges at third order.

The differences between the waveforms at $\scri$ for resolution
$h=0.64M$ are of order of $0.03\%$ in amplitude with a dephasing
of $0.002$ radians over the course of the evolution.

It should be noted that the differences in the waves between the two
world-tube locations is one order of magnitude smaller than the
numerical errors in amplitude and phase inherent in the Cauchy evolution~\cite{Pollney:2009yz}.
We therefore expect that the dominant error in the waveforms is due to the discretisation error during Cauchy evolution.
Indeed, using the three resolutions $h=0.96M$, $h=0.80M$ and $h=0.64M$, we
find a very similar convergence order for (a) CCE with fixed world-tube
location, and (b) for a pure Cauchy evolution with the same fixed finite radius
extraction (compare \cite{Pollney:2009yz}).

However, it should also be noted that the (non-convergent) differences between extrapolated waveforms and waveforms obtained at $\scri$ as found in \cite{Reisswig:2009us}
are of the same order as the (convergent) discretisation error inherent in the Cauchy evolution for our given resolutions
as found in \cite{Pollney:2009ut}. Further increasing the resolution of the Cauchy evolution will therefore lead to an error that is much smaller than the error
made by finite radius extrapolation.
This highlights the importance of the systematic error due to
extrapolation
from finite radius, which is removed by performing computations at $\scri$ via CCE.

\subsection{Conservation of energy and angular momentum}

We have used the measured waveforms to compute the initial mass and
angular momentum of the modeled spacetimes.
The initial mass is given by the ADM mass as computed at infinity on the initial slice using the methods described in \cite{Ansorg:2004ds}.
Since the initial slice is conformally flat, the initial angular momentum of the spacetime is simply given by (see
for example~\cite{Cook94, Cook00a, Pfeiffer:2000um})
\begin{equation}
  \mathbf{J}_{_{\mathrm{ADM}}}
    = \mathbf{C}_1 \times \mathbf{p}_1 +
    \mathbf{C}_2 \times \mathbf{p}_2 + \mathbf{S}_1 + \mathbf{S}_2 \ .
  \label{eq:InitialSpin}
\end{equation}
Here $\mathbf{C}_i$, $\mathbf{p}_i$ and $\mathbf{S}_i$ are the position, the
linear momentum and the spin of the $i$-th black hole.

After evolution, \ie, after the horizon of the final black hole has equilibriated, we compute the mass and spin
of the remnant on its apparent horizon using the isolated horizon formalism~\cite{Dreyer02a, ashtekar03a, Schnetter-Krishnan-Beyer-2006}, or alternatively, 
by using the horizon circumference formula \cite{Alcubierre:2004hr} (the two methods agree within the reported errors as given in Table~\ref{tab:binaries}, where we use
the values obtained from the circumference formula\footnote{The close agreement between both methods further supports the idea that the final black hole is indeed a Kerr black hole.}).
In addition, we compute the radiated energy and angular momentum carried by the gravitational waves
using the results of finite radius computations, from extraction at $\scri$, and
from an extrapolation of the results at several finite radii.
The finite radius computations are obtained at an extraction world-tube $r_{\Gamma}=100M$, which is typical for current
numerical relativity simulations (see \eg,~\cite{Hannam:2009hh, Aylott:2009ya}).

\begin{table}
  \begin{ruledtabular}
    \begin{tabular}{lcc}
      Model                                            &  $(a_1,a_2)=(0,0)$                 &  $(a_1,a_2)=(0.8, -0.8)$ \\ \hline
      total ADM mass $M_{\rm ADM}$                     &  $0.99051968 \pm 2\times 10^{-8}$  &  $0.98826246 \pm 2\times 10^{-8}$ \\
      initial angular momentum $J_{\rm ADM}$           &  $0.993300 \pm 1.0\times 10^{-16}$ &  $0.900312 \pm 1.0\times 10^{-16}$ \\
      irr.~Mass $M_{\rm f,irr}$                        &  $0.884355 \pm 2.0\times 10^{-5}$  &  $0.883636 \pm 1.0\times10^{-4}$      \\
      Spin $S_f/M_f^2$                                 &  $0.686923 \pm 1.0\times 10^{-5}$  &  $0.686300 \pm 2.7\times 10^{-4}$        \\ 
      Mass $M_{f}$                                     &  $0.951764 \pm 2.0\times 10^{-5}$  &  $0.950829 \pm 3.8\times 10^{-5}$  \\ 
      Angular momentum $S_f$                           &  $0.622252 \pm 1.0\times 10^{-5}$  &  $0.620469 \pm 1.9\times 10^{-4}$ \\ \hline
      Recoil velocity $v_{\rm kick,CCE}$               &  --                                & $364.30 \pm 0.32\, {\rm km/s}$ \\ 
      Radiated angular momentum $J_{\rm rad,CCE}$            &  $0.370704 \pm 6.2\times10^{-5}$    & $0.286908 \pm 2.6\times10^{-4}$  \\
      Radiated energy $E_{\rm rad,CCE}$                      &  $0.038828 \pm 2.5\times10^{-6}$    & $0.037689 \pm 4.3\times10^{-5}$  \\ \hline
      Recoil velocity $v_{\rm kick,finite,r=100}$            &  --                                 & $350.37 \pm 1.46\, {\rm km/s}$ \\
      Radiated angular momentum $J_{\rm rad, finite,r=100}$  &  $0.339725 \pm 4.0\times10^{-4}$    & $0.266654 \pm 5.7\times10^{-3}$  \\
      Radiated energy $E_{\rm rad, finite,r=100}$            &  $0.037354 \pm 3.9\times10^{-5}$    & $0.035369 \pm 3.7\times10^{-4}$\\ \hline
      Recoil velocity $v_{\rm kick,extrap}$                  &  --                                 & $362.87 \pm 1.19\, {\rm km/s}$ \\
      Radiated angular momentum $J_{\rm rad,extrap}$         &  $0.370007 \pm 6.8\times10^{-5}$    & $0.283738 \pm 2.1 \times 10^{-3}$  \\
      Radiated energy $E_{\rm rad,extrap}$                   &  $0.038715 \pm 2.2\times10^{-6}$    & $0.037189 \pm 2.2\times10^{-4}$ \\ \hline
      Residual $|v_{\rm kick,CCE}-v_{\rm kick,extrap}|$      &  --                                 & $1.43 {\rm km/s}$ \\ 
      Residual $|J_{\rm rad,CCE}-J_{\rm rad,extrap}|$        &  $7.0\times10^{-4}$                 & $3.1\times 10^{-3}$ \\
      Residual $|E_{\rm rad,CCE}-E_{\rm rad,extrap}|$        &  $1.1\times10^{-4}$                 & $5.0\times10^{-4}$  \\ \hline
    \end{tabular}
  \end{ruledtabular}
  \caption{\label{tab:binaries}
Properties of the initial spacetime, of the merger remnant, and
of the emitted radiation calculated from the appropriate waveform. 
The reported errors of the extrapolated values refer to the error in
the extrapolation and the errors of the values at $\scri$ refer to the error from two different
extraction world-tubes at the same fixed resolution. 
The error in the initial angular momentum $J_{\rm ADM}$ is at machine precision as it is known analytically
in terms of the initial parameters of the black holes.
All other errors are given by the Richardson expanded differences between a medium
and a high resolution run as described in \cite{Pollney:2009yz}.
The last three rows display the differences between extrapolated results and results at $\scri$, 
which we note is of the order of the discretisation error as indicated by the finite radius error-bars.
   }
\end{table}

The radiated energy and angular momentum flux can be calculated in terms of spin-weighted spherical harmonic coefficients
of $\psi_4$ as derived in \cite{Lousto:2007mh, Ruiz:2008multipole}.
The radiated energy flux is given by
\be
\frac{dE}{dt}
= \lim_{r\rightarrow\infty} \left\{ \frac{r^2}{16\pi} \sum_{\ell,m} \left| \int_{-\infty}^{t}d\tilde{t} \psi_4^{\ell m} \right|^2 \right\}
\label{Erad}
\ee
and the components of the radiated angular momentum flux read
\begin{widetext}
\bea
\frac{dJ_x}{dt} &=& \lim_{r\rightarrow\infty} \left\{ \frac{r^2}{32\pi} \Im \sum_{\ell,m} \int_{-\infty}^{t}dt'\int_{-\infty}^{t'}dt'' \psi_4^{\ell, m} \times \int_{-\infty}^{t}dt' \left( f_{\ell, m}\bar{\psi}_4^{\ell, m+1} + f_{l,-m}\bar{\psi}_4^{\ell, m-1} \right) \right\}, \label{eq:Jx}\\
\frac{dJ_y}{dt} &=& \lim_{r\rightarrow\infty} \left\{ \frac{r^2}{32\pi} \Re \sum_{\ell,m} \int_{-\infty}^{t}dt'\int_{-\infty}^{t'}dt'' \psi_4^{\ell, m} \times \int_{-\infty}^{t}dt' \left( f_{\ell, m}\bar{\psi}_4^{\ell, m+1} - f_{l,-m}\bar{\psi}_4^{\ell, m-1} \right) \right\}, \label{eq:Jy}\\
\frac{dJ_z}{dt} &=& \lim_{r\rightarrow\infty} \left\{ \frac{r^2}{16\pi} \Im \sum_{\ell,m} \int_{-\infty}^{t}dt'\int_{-\infty}^{t'}dt'' \psi_4^{\ell, m} \times \int_{-\infty}^{t}dt' \bar{\psi}_4^{\ell, m} \right\}, \label{eq:Jz}
\eea
\end{widetext}
where 
\be
f_{\ell, m} = \sqrt{\ell(\ell+1)-m(m+1)}.
\ee

Note that we can also compute the expressions for radiated energy and angular momentum in terms of the news function $N$ by using
the fact that in a Bondi-frame, $\Psi_4$ is related to the news via
\be
\Psi_4 = \p_u N\,.
\ee
Hence, we can replace all first time integrals of $\psi_4$ in (\ref{Erad})--(\ref{eq:Jz}) by $N$.

We find that modes are well resolved only up to $(\ell,m)=(8,8)$,
and hence the sum in the expression above is to $\ell=8$.
We have computed the sums for waveforms extracted at finite radii and at $\scri$.
The finite radius computations have been performed on extraction spheres
$r=\{280M,300M,400M,500M,600M,1000M\}$ for the non-spinning binary, and $r=\{100M,250M,500M\}$
for the spinning binary.
Given the radiated energy and angular momentum on the three extraction spheres, we do a linear extrapolation by fitting a function of the form
\be
Q(r) = Q_0 + Q_1/r\,, \label{eq:fit}
\ee
where $Q$ is either $E_{\rm rad}$ or $J_{\rm rad}$.

The values for the initial mass $M_{\rm ADM}$, initial angular momentum $J_{\rm ADM}$, final (Christodoulou) mass $M_{f}$, final angular momentum $S_f$, as well as radiated energy $E_{\rm rad}$
and radiated angular momentum $J_{\rm rad}$ are reported in Table~\ref{tab:binaries} for both binary systems.
The reported errors refer to Richardson expanded differences between a high and a medium resolution run as described in \cite{Pollney:2009yz}, except for
the error in the extrapolation and the error in CCE, where for the former we report the error in the fit, and for the latter report the differences between two given extraction world-tubes for the
same fixed resolution. Hence, they do not contain the discretisation error of the evolution, which, in all cases, is one order of magnitude larger than the error in the extrapolation and in CCE.

Table~\ref{tab:binaries} lists the radiated quantities determined from
the measured values of $\psi_4$. However, since we also determine the news function $N$ at $\scri$, we can independently
compute $E_{\rm rad, CCE}$ and $J_{\rm rad,CCE}$ based on $N$ in order to cross check the result. We find that the differences in all cases are on the order of the reported error bars, hence further 
validating the results based on CCE.

Another important point to make is that the differences between extrapolated data and data obtained at $\scri$ are of the same order of magnitude as the discretisation error in the Cauchy
evolution. This is indicated by the last three rows of Table~\ref{tab:binaries}. The discretisation error of these quantities due to the the Cauchy evolution is given by the error-bars
of the finite radius quantities.

Due to conservation of energy and angular momentum, the following relations must hold exactly
\be
S_f+J_{\rm rad} = J_{\rm ADM}\,, \qquad M_f+E_{\rm rad} = M_{\rm ADM}\,.
\label{eq:conservation}
\ee
The residuals of these relations are reported in Table~\ref{tab:conservation} for the quantities at a finite radius, extrapolated to infinity,
and computed at $\scri$, for both binary systems.

Note that the residuals in the extrapolated values of the two binaries are of
comparable accuracy to the values obtained via CCE. This is surprising, 
considering the error in the extrapolation procedure itself, and
demonstrates that extrapolation, even in the most extreme spinning cases, 
delivers comparable performance. Note, however, that our extrapolation uses 
large extraction radii compared to common practice in numerical relativity. By using a set of smaller extraction radii in the 
extrapolation, we expect that the errors increase up to one order of magnitude
(compare \cite{Pollney:2009ut}). In addition, the systematic error
made in the extrapolation will not converge with numerical resolution, but only linearly with extraction radius $\mathcal{O}(r)$.
This however, puts high computational demands on future high accuracy simulations.

In all cases, we observe that a pure finite radius computation based on an extraction sphere
at $r=100M$ clearly performs worse and has an error that is larger
by 1--2 orders of magnitude, which, in the
case of angular momentum, corresponds to an error of the order of $1.0\%-3.0\%$.

\begin{table}
  \begin{ruledtabular}
    \begin{tabular}{lcc}
      Model                                           &  $(a_1,a_2)=(0,0)$           & $(a_1,a_2)=(0.8,-0.8)$ \\ \hline
      $|S_f+J_{\rm rad,finite}-J_{\rm ADM}|$           & $3.1 \times 10^{-2}$         & $1.3\times10^{-2}$ \\
      $|S_f+J_{\rm rad,extrap}-J_{\rm ADM}|$           & $1.0 \times 10^{-3}$         & $3.9\times10^{-3}$ \\            
      $|S_f+J_{\rm rad,CCE}-J_{\rm ADM}|$                  & $3.4\times 10^{-4}$          & $7.1\times10^{-3}$ \\ \hline
      $|M_f+E_{\rm rad,finite}-M_{\rm ADM}|$           & $1.4 \times 10^{-3}$         & $2.1\times10^{-3}$ \\
      $|M_f+E_{\rm rad,extrap}-M_{\rm ADM}|$           & $4.1\times 10^{-5}$          & $2.4\times10^{-4}$ \\
      $|M_f+E_{\rm rad,CCE}-M_{\rm ADM}|$                  & $7.2\times 10^{-5}$          & $2.6\times10^{-4}$ \\
    \end{tabular}
  \end{ruledtabular}
  \caption{\label{tab:conservation}
           Residuals in conservation of energy and angular momentum (\ref{eq:conservation}) for the non-spinning and 
           spinning configuration and for radiated mass $E_{\rm rad}$ and angular momentum $J_{\rm rad}$
           as calculate at a finite radius $r=100M$, extrapolated from finite radii computations to infinity, and by the CCE computation at $\scri$.
           Both, extrapolated quantities and those directly computed at $\scri$ satisfy the conservation conditions with about the same amount of accuracy.
           Not using any extrapolation but relying on a pure finite radius computation leads to an error that is of about 1--2 orders of magnitude larger than 
           the one found by the other two methods.
   }
\end{table}

\subsection{Recoil velocity}

In addition to energy and angular momentum, gravitational waves also carry linear momentum.
Depending on the binary configuration, there might be a prefered direction at which radiation, and hence linear momentum, is beamed.
Due to conservation of linear momentum, this gives rise to a
gravitational recoil, or ``kick'' acquired by the merger remnant.
An accurate measurement of this is effect is crucial in various fields of astrophysics as this determines whether the final black hole is kicked out
of its host environment. Clearly, the existence or non-existence of a black hole can have a dramatic impact on the host and determines its further evolution.

Previous studies have already considered this effect in detail, and particularly, the binary configuration that we consider here
belongs to a sequence that has already been examined in \cite{Herrmann:2007ac, Rezzolla-etal-2007}.
However, these studies estimated the recoil velocity using waveforms extracted at a finite radius.
In this subsection, we measure this effect where it is defined unambiguously, namely at $\scri$.

The radiated linear momentum flux can be calculated in terms of spin-weighted spherical harmonic coefficients
as derived in \cite{Ruiz:2008multipole} and given by
\begin{widetext}
\bea
\frac{dP_x}{dt}+i\frac{dP_y}{dt} &=& \lim_{r\rightarrow\infty} \left\{ \frac{r^2}{8\pi} \sum_{\ell,m} \int_{-\infty}^{t}dt'\psi_4^{\ell, m} \nonumber \right. \\
   & & \qquad \left. \times \int_{-\infty}^{t}dt' \left( a_{\ell, m}\bar{\psi}_4^{\ell, m+1} + b_{l,-m}\bar{\psi}_4^{\ell-1, m+1} - b_{l+1,m+1}\bar{\psi}_4^{\ell+1, m+1} \right) \right\}, \\
\frac{dP_z}{dt} &=& \lim_{r\rightarrow\infty} \left\{ \frac{r^2}{16\pi} \sum_{\ell,m} \int_{-\infty}^{t}dt'\psi_4^{\ell, m} \right. \nonumber \\
   &  & \left. \qquad \times \int_{-\infty}^{t}dt' \left( c_{\ell,m}\bar{\psi}_4^{\ell, m} + d_{\ell,m}\bar{\psi}_4^{\ell-1, m} + d_{\ell+1,m}\bar{\psi}_4^{\ell+1, m}\right) \right\},
\eea
\end{widetext}
where 
\bea
a_{\ell, m} &=& \frac{\sqrt{(\ell-m)(\ell+m+1)}}{\ell(\ell+1)}, \\
b_{\ell, m} &=& \frac{1}{2\ell}\sqrt{\frac{(\ell-2)(\ell+2)(\ell+m)(\ell+m-1)}{(2\ell-1)(2\ell+1)}}, \\
c_{\ell, m} &=& \frac{2m}{\ell(\ell+1)}, \\
d_{\ell, m} &=& \frac{1}{\ell}\sqrt{\frac{(\ell-2)(\ell+2)(\ell-m)(\ell+m)}{(2\ell-1)(2\ell+1)}}.
\eea
Again, we have computed the expression above by including modes up to
$\ell\leq8$. In addition, we calculate this effect using
a finite radius computation and a computation at $\scri$. Given the three extraction spheres at $r=\{100M,250M,500M\}$ for the spinning configuration, we have extrapolated
the recoil velocity to infinity by means of Eq.~(\ref{eq:fit}).

It should be noted that it is crucial to include the appropriate vector integration constant when integrating the linear momentum flux in time.
As analyzed in \cite{Pollney:2007ss}, there are two important contributions. The first mainly comes from the initial burst of junk radiation
that adds significantly to the measured recoil at the beginning of the simulation. The second is due to the non-zero net-linear momentum that the binary had acquired if it had 
inspiralled from an infinite separation.
We have measured the correct vector integration constant by applying the method described in \cite{Pollney:2007ss} and corrected the computation accordingly.
We find good agreement between finite radius computations, extrapolation to infinity, and computation at $\scri$.
The results are included in Table \ref{tab:binaries}.

\subsection{Constraint preservation}

We check the preservation of the constraints at the world-tube and at $\scri$ for different radii and resolutions.
We use the constraint equations as implemented and first tested with linearized solutions in \cite{Reisswig:2006}.
As a consequence of the Bondi system, there are three constraint equations
\be
R_{00}=0\,,\qquad R_{01}=0\,, \qquad q^AR_{0A}=0\,.
\ee
Analytically, if these constraints are satisfied initially and at the world-tube, they should be satisfied
exactly everywhere and at all times. However, in computer simulations, the solution always contains numerical error so that
the constraints can only be satisfied approximately. Since this error is only due to discretisation, it should converge
away in the limit of infinite resolution. Indeed, this is what we find.

\begin{figure}
  \centering
  \includegraphics[width=8.0cm]{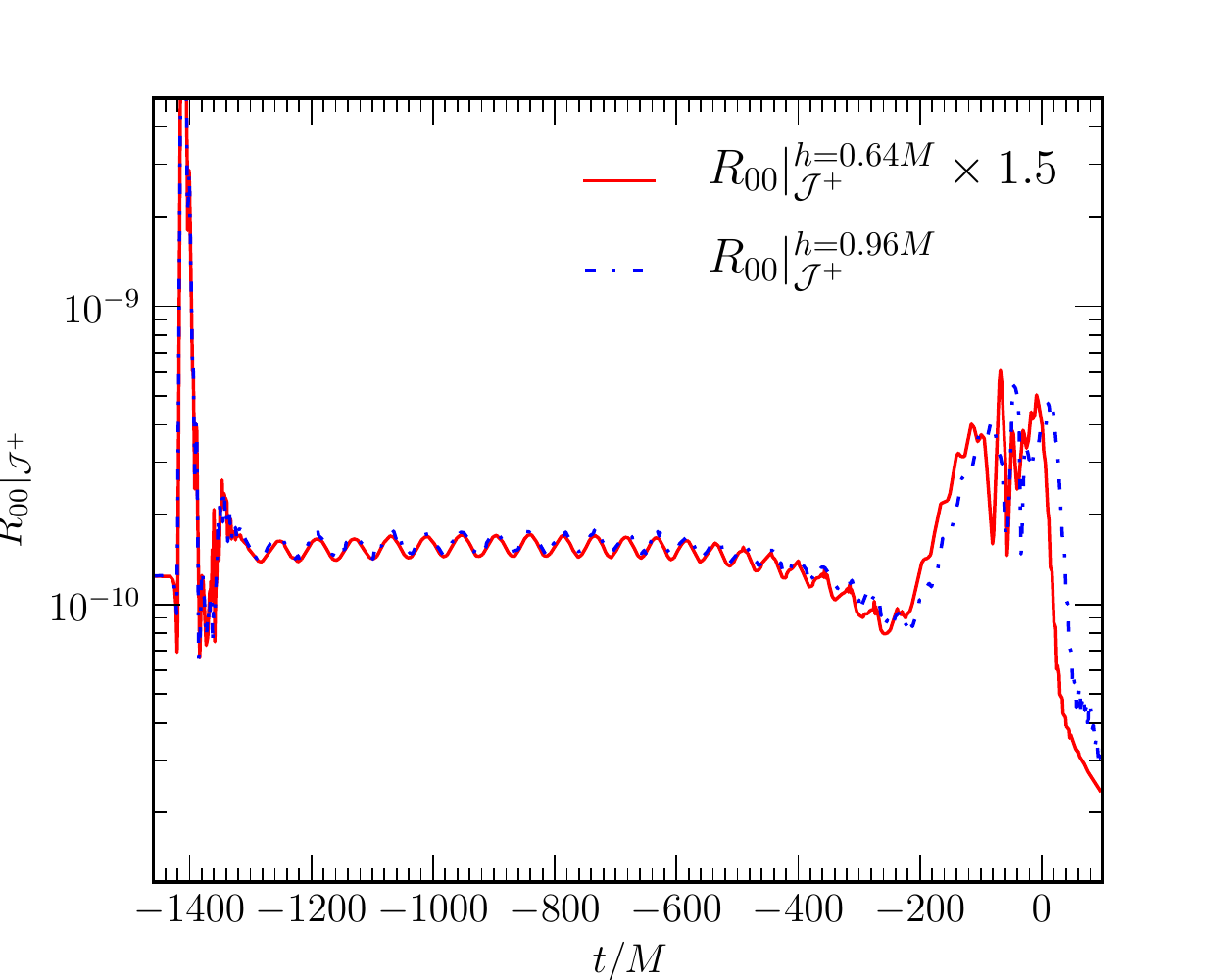}\hfill
  \includegraphics[width=8.0cm]{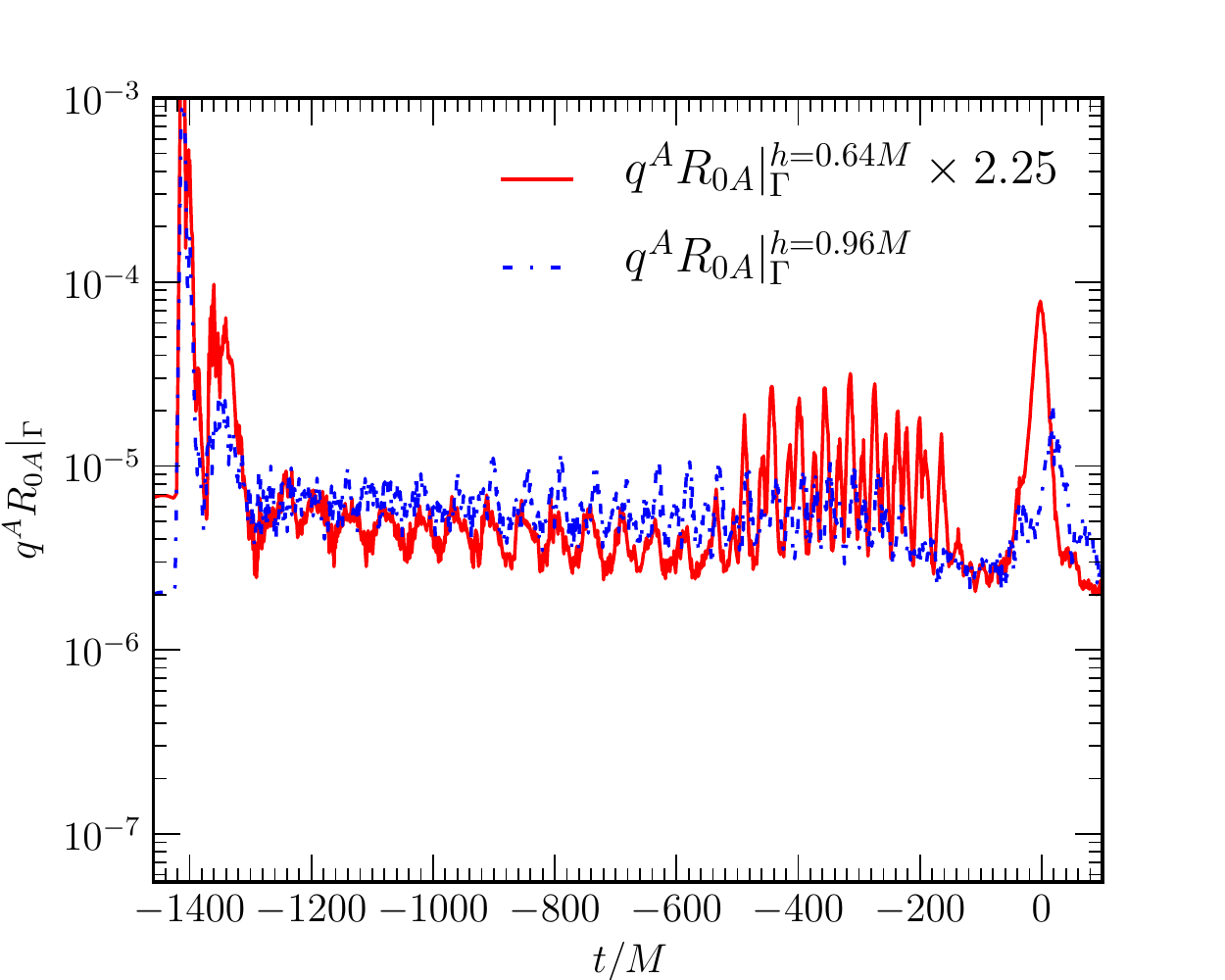}\hfill
  \includegraphics[width=8.0cm]{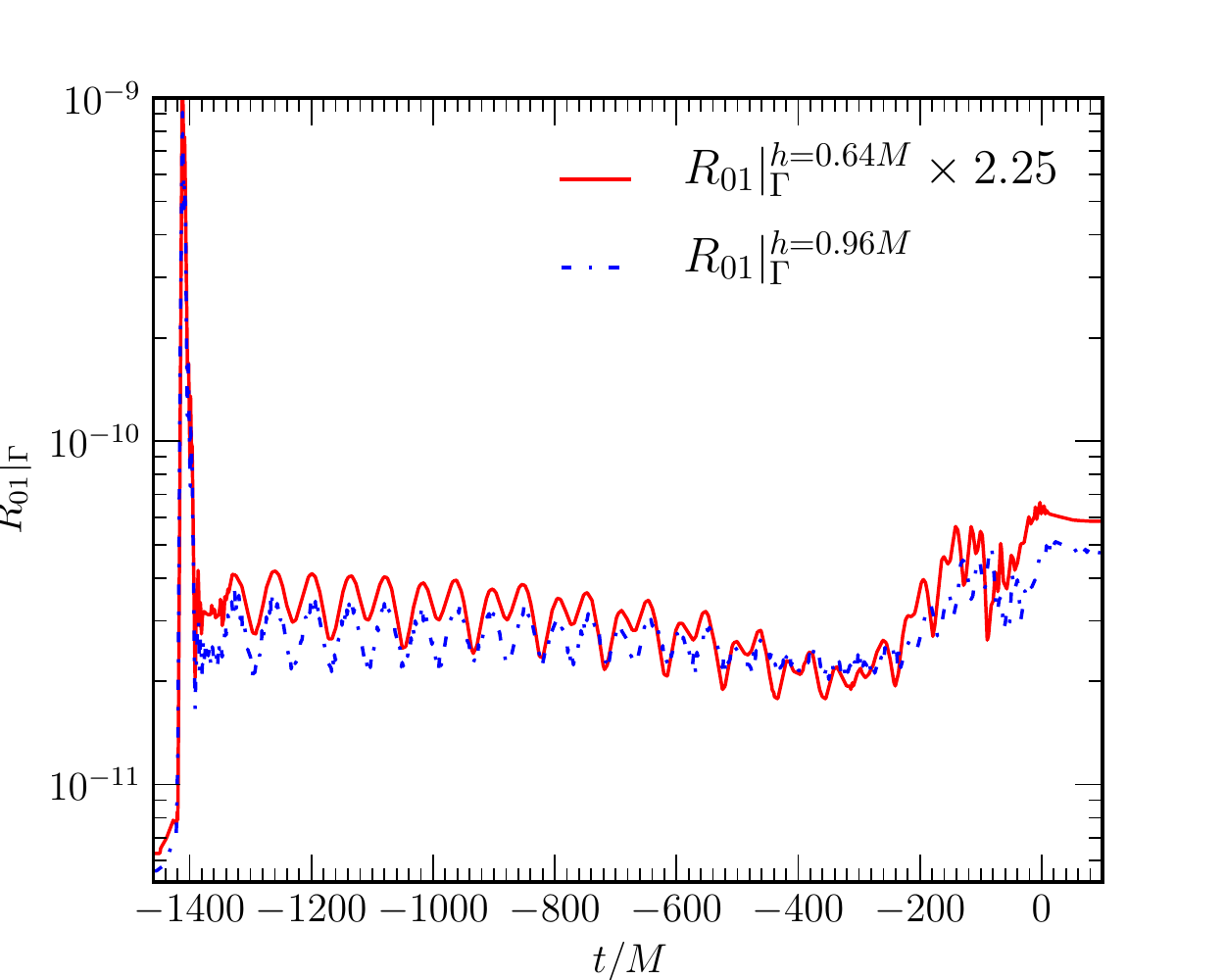}\hfill
  \caption{The \textit{left} panel shows the $L_2$-norm of the constraint $R_{00}$ at $\scri$ for the two resolutions $h=0.96M$ and $h=0.64M$ scaled for first-order convergence.
           The \textit{right} figure shows the $L_2$-norm of the constraint $q^AR_{0A}$ at the world-tube $\Gamma$ for the two resolutions $h=0.96M$ and $h=0.64M$ scaled for second-order convergence.
           The \textit{bottom} panel shows the same for the constraint $R_{01}$ at the world-tube $\Gamma$.
           All figures refer to the non-spinning binary.}
    \label{fig:constraints}
\end{figure}

Fig.~\ref{fig:constraints} shows the convergence of $R_{00}$ at $\scri$ and $q^AR_{0A}$ and $R_{01}$ at the world-tube $\Gamma$ for the non-spinning binary.
While $R_{00}$ appears to be only first-order convergent at $\scri$, we note that the error is already
very small and at the order of $10^{-10}$ and hence close to machine precision.
The same applies to the constraint $R_{01}$ which exhibits a convergence exponent between first and second-order.
The constraint $q^AR_{0A}$, on the other hand, exhibits an overall second-order convergence exponent at the world-tube.
The origin of the spurious spikes occuring at the high-resolution run between $-500<t<-100$ and reducing the convergence order, is currently unkown 
but may be due to the first-order knowledge of $U_{,r}$ at the world-tube. This is subject to future investigation and improvements of the algorithm and code.
We also note that during merger, the convergence of the $q^AR_{0A}$ constraint is reduced at the world-tube. This, again, may be due to the first-order knowledge 
of $U_{,r}$ at the world-tube. However, the convergence of the constraints appeared to be problematic already in \cite{Reisswig:2006} and we emphasize that
the differences in the waveforms at $\scri$ as extracted from different world-tube locations converge to zero at approximately second-order and independent of the constraints. 

In all cases, the constraints exhibit a large spike at the time when the first non-trivial data, particularly the initial burst of junk radiation, reaches the world-tube location.
Although the junk radiation should satisfy the constraint equations since it arises as part of the solution to the constraint preserving initial data at the initial Cauchy hypersurface,
we note that there is principally an incompatibility between the conformal flatness assumption of the 3-metric on the initial Cauchy hypersurface and the initial data
of the 4-metric, $J(u=0)=0$, on the initial null hypersurface. This may lead to the observed initial spikes in the constraints.
Further investigation is necessary to clearly identify the source of error and to improve the construction of characteristic initial data.

\section{Notes on characteristic code parameters}
\label{s-param}

There are some subtleties regarding the choice of some code and grid parameters that
have been found empirically during tests of the extraction method.
Here, we collect some of these findings for future reference.

\begin{itemize}
\item
We use conformally flat initial data on the initial null hypersurface, \ie $J(u=0)=0$.
This assumption is in principle incompatible with the proper radiation content of a binary black hole spacetime, and it
is subject to the same error as on the Cauchy side, where we use conformally flat initial data.
In practice, the spurious burst of junk radiation at the beginning of the simulation quickly leaves the system
and does not affect the radiation at the accuracy given by the numerics.

\item
The radial resolution of the physical coordinate on the characteristic grid should be at least as fine at the world-tube as
the radial resolution on the Cauchy side.
We can estimate the compactified radial resolution in terms of the physical resolution via
\be
\frac{dr}{dx}=r_{\rm wt}\left(\frac{1}{1-x}+\frac{x}{(1-x)^2} \right)\,,
\ee
which asymptotes to $r_{\rm wt}$ for $x\rightarrow 0$ at the world-tube, \ie
\be
x\rightarrow 0, \qquad \frac{dr}{dx}\rightarrow r_{\rm wt}\,.
\ee
It follows that
\be
\Delta r\vert_{\Gamma} \sim r_{\rm wt} \Delta x
\ee
at the world-tube.
We can then select an appropriate $\Delta x$, so that $\Delta r\vert_{\Gamma}$ at the world-tube is at least as small as the radial resolution $\Delta r$ on the Cauchy side.

It turns out that at least $N_x\approx200$ radial points are required to resolve the wave properly at $\scri$. 
Since $\Delta R=0.64M$ at the extraction spheres for the highest resolution, the compactification parameter has to be
$r_{\rm wt}\approx 256$ in order to meet the condition $\Delta r\leq\Delta R$.
Further increasing the characteristic radial resolution while leaving the Cauchy $\Delta R=\textrm{const.}$ has no noticeable influence on the wave at $\scri$.

\item
The linearized approximation for the conformal factor $\omega$ is used for the computation of the news $N$.
The full non-linear computation results in spurious drifts and does not converge when considering the Schwarzschild test. This is potentially a bug in the existing implementation
and affects only the news. $\Psi_4$ is only affected at higher-order which does not play a role in practice. However, as the fields at $\scri$ are in the linear regime, the linear approximation 
is a valid simplification.

\item
We use the first-order evolution system in the characteristic code, \ie, we use 
only first eth-derivatives. The implementation with second eth-derivatives
does not converge and gives inconsistent results for the binary waveform when 
considering different world-tube locations.
This might also be related to the problem formulated in \cite{Gomez01, Gomez02a}.
In principle, both methods should be equivalent and the non-convergence of the 
second-order system can be attributed to a bug in the code.

\item
The news and $\Psi_4$ are computed using the second-order implementation of the 
extraction module, \ie, using second eth-derivatives.

\item
All quantities are decomposed in terms of the complex-valued spin-weighted 
spherical harmonics ${}_sY_{\ell m}$.

\end{itemize}

\section{Conclusions}
\label{s-conc}

We have demonstrated results from an implementation of characteristic
extraction in numerical relativity, applied to the astrophysically
relevant test cases of binary black holes. Our implementation
carries out the characteristic extraction as a post-processing step.
As such, the code is general
purpose and can be applied to boundary data generated by any Cauchy
simulation, independent of numerical scheme or formulation of the
Einstein equations. The code has been tested against analytic solutions, and it
was found that it exhibits the expected convergence properties, as
well as gauge-independence, a property lacking in finite radius calculations.
Applied to two test cases involving the inspiral and merger
of binary black holes, we have determined the emitted gravitational radiation, and
found that the CCE method does indeed generate results which are
independent of the world-tube on which the data has been extracted.
We have also determined
the residuals in the conservation of mass and angular momentum, as well as
the recoil velocity (when the black hole spins are
anti-aligned). We find
that the residuals of the CCE method are very similar to those obtained by extrapolation
to infinity of results obtained using a pure
Cauchy evolution, providing an important confirmation of the usefulness
of such methods. However, the (non-convergent) differences between both measures is already
on the order of the discretisation error of the Cauchy evolution, therefore indicating
that as the resolution is further increased, the finite radius extrapolation will be the
dominant source of error in the wave-extraction. The error in CCE, 
as measured here for a particular set of resolutions is one order of magnitude smaller,
and will converge away if the resolution is further increased.

Characteristic extraction yields results that are free of a number of
systematic errors in the modeling
of the physical problem. Placing the outer boundary at a sufficiently large
distance, we are able to causally disconnect the artificial boundary
from
the extraction world-tube.
However, there is still the possibility of a systematic error in the extent
to which there is an unphysical radiation content in the initial data. Further,
it should be noted that there is, in general, an inconsistency between the
conformally flat initial data used in the Cauchy computation, and that of
$J=0$ used on the initial characteristic null cone. It is usually assumed
that the spurious initial radiation will quickly be flushed out of the
system in a burst of initial junk radiation. The observed convergence of
the differences between the radiation signals at different extraction radii,
indicates that, after the burst of junk radiation,
the effect of spurious, unphysical initial data is very small and below the 
discretisation error.

Future developments will clearly include applying characteristic
extraction to a variety of other physical systems. It will also be
interesting to revisit the problem of Cauchy-characteristic matching,
in which the extraction world-tube and the Cauchy outer boundary
coincide and boundary information is passed in both directions during
the course of the evolution. In this case evolution in the Cauchy
domain between, say $r=100M$ and $r=3000M$, is avoided so the method
is significantly more efficient; and further one can avoid inconsistencies in
the initial data between the two domains. Finally, it would be useful
to construct initial data that correctly models the physical problem;
this is, in principle, possible at least for the characteristic
domain.

\acknowledgments

We thank Stanislav Babak, Ian Hinder, Luciano Rezzolla for useful discussions.
CR and DP thank Rhodes University, and NTB thanks
Max-Planck-Institut f\" ur Gravitationsphysik, for hospitality.
This work was supported by international collaboration grants funded by
National Research Foundation, South Africa, Bundesministerium f\"ur Bildung und
Forschung, Germany, and DFG grant
SFB/Transregio~7 ``Gravitational Wave Astronomy''.
DP was supported by a grant from the VESF.
BS was supported by grants from the Sherman Fairchild Foundation, by NSF grants
DMS-0553302, PHY-0601459, PHY-0652995, and by NASA grant NNX09AF97G.
Computations were performed at the AEI, at LRZ-M\"unchen, on Teragrid clusters
(allocation TG-MCA02N014), and LONI resources at LSU.

\bibliography{aeireferences.bib}

\end{document}